\documentclass{aa}
\usepackage{graphics}
\usepackage{epsfig}

\begin{document}

   \title{Near--IR observations of NGC 6822: AGB stars, distance, metallicity and structure}

   \author{M.-R.L. Cioni\inst{1}
          \and
           H.J. Habing\inst{2}
          }

   \offprints{mcioni@eso.org}

   \institute{European Southern Observatory,
              Karl--Schwarzschild--Stra\ss e 2,
              D--85748 Garching bei M\"{u}nchen, Germany
              \and
              Sterrewacht Leiden,
              Niels Bohrweg 2,
              2333 RA Leiden, The Netherlands
             }

   \date{Received 7 July 2004/ Accepted 8 September 2004}

   \titlerunning{Red giants in NGC 6822}

   \abstract{  Observations in  the  $IJK_s$ wave  bands covering  the
central $20^{\prime}\times 20^{\prime}$ of  the Local Group galaxy NGC
6822  have been  made with  the William  Herschel Telescope  in La
Palma.  They  have allowed us to  characterize, for the  first time in
the  near--infrared across  the whole  galaxy, its  late--type stellar
population (i.e.  red giant and  asymptotic giant branch stars) and to
derive from the ratio between carbon--rich and oxygen--rich asymptotic
giant  branch   stars  an  indication  about   spatial  variations  in
metallicity. These amount to about  $1.56$ dex, twice of what has been
previously  found   within  each  Magellanic  Cloud   using  the  same
technique. We have calibrated  our photometry on the DENIS ($I$--band)
and 2MASS ($J$ and $K_{\mathrm s}$ bands) data and obtained a distance
modulus of $(m-M)_0=23.34\pm0.12$ from the  position of the tip of the
RGB. The  large scale distribution  of late--type stars  suggests that
either  the galaxy is viewed under a high inclination angle or  it has  a
non--negligible  thickness.    
}

\maketitle

\section{Introduction}

NGC 6822 (Barnard \cite{bar}) is an isolated irregular galaxy of the
Local Group, the nearest to the Milky Way after the Sagittarius Dwarf
and the Magellanic Clouds. Its stellar population has been largely
studied through observations in the optical wave bands (Gallart et al.
\cite{galla1}, \cite{galla2}, \cite{galla94}, Komiyama et al.
\cite{komi}, Letarte et al. \cite{tarte}, Wyder \cite{wyde}, de Blok
\& Walter \cite{blok3}).  Similar to the Magellanic Clouds the galaxy
contains a large and widely distributed intermediate--age stellar
population ($14^\prime \times 18^\prime$) and a relatively small
optical bar ($6^\prime \times 11^\prime$). NGC 6822 is embedded in a
much larger ($42^\prime \times 19^\prime$) HI envelope (cft. Volders
\& H\"{o}gbom \cite{vold}, Roberts \cite{rob}, de Blok \& Walter
\cite{blok}, Weldrake et al. \cite{weld}) and because of its low
galactic latitude ($b=-18.39^\circ$) it is affected by a moderate
foreground extinction (Schlegel et al. \cite{sch}) and contaminated by
foreground stars. Schlegel et al. (\cite{sch}) derived a mean reddening of
$E(B-V)=0.25$ in agreement with measurements by Massey et al.
(\cite{mass}) in the outer eastern and western part of the galaxy.
The latter measured $E(B-V)=0.45$ near the center of the galaxy, but
this may be due to dust embedding young objects; this amount of
reddening hardly affects the photometry of red giant branch stars
(RGB) or asymptotic giant branch (AGB) stars. The mean distance to the
galaxy is $497$ kpc, or $(m-M)_0=23.48\pm0.08$ (Van den Berg
\cite{vdb}).

An abundance gradient in [O/H] has been suggested by Venn et al.
(\cite{venn}); their measurements agree with those in HII
regions (Pagel et al. \cite{pagel}, Skillman et al. \cite{skill},
Chandar et al. \cite{chan}). The mean [O/H] abundance is intermediate
between that of the SMC and LMC. Tolstoy et al. (\cite{tolto}) derived the
metallicity [Fe/H] of a few field RGB stars from measurements of the
Ca II triplet. They obtained on average [Fe/H]$=-0.9$ dex but the
individual points spread over a range from $-2.0$ to $-0.5$ dex.

NGC 6822 started to form stars at least $10$ Gyr ago and from low
metallicity gas as the detection of RR Lyrae stars shows (Baldacci et
al. \cite{balda}, Clementini et al. \cite{clem}). 
Gallart et al. (\cite{galla2}) and Wyder
(\cite{wyde}), from their synthetic analysis of the RGB and of the
whole HST colour--magnitude diagram, respectively, derived an age of
$12$--$15$ Gyr for the oldest stars.  About 3 Gyr ago the rate of
star formation began to increase (Tolstoy et al.  \cite{tolto}) and
it kept increasing over the last $100-200$ Myr (Gallart et al.
\cite{galla3}, Hutchings et al. \cite{hutch}). Blue stars that formed
in the past $0.6$ Gyr are not well mixed with the rest of the galaxy
(Wyder \cite{wyde}) and trace the distribution of the HI gas (Komiyama
et al. \cite{komi}).  Many variable stars were found at the tip of the
RGB that are probably long period variables (Baldacci et al.
\cite{baldav}) and thus AGB stars. These AGB stars are $1$--$10$ Gyr
old and have Z$=0.001$--$0.004$ (Gallart et al.  \cite{galla2}).

The only published near--infrared ($J$ \& $K_{\mathrm s}$ bands)
photometry of late--type stars is that by Davidge (\cite{dav}), except
for about $20$ known red-supergiants observed and discussed by Elias
\& Frogel (\cite{eli}).  Davidge observed three small
($0.5^{\prime\prime}\times 0.5^{\prime\prime}$) fields reaching
$K=21$.  His analysis suggests that both age and metallicity can be
disentangled from the slope and zero-point of the giant branch in the
($J-K_{\mathrm s}$, $K_{\mathrm s}$) diagram.  In fact he concludes
that the slope of the RGB indicates [Fe/H]$=-1.0\pm0.3$ by comparison
with theoretical isochrones (Girardi et al.  \cite{leo}), while the
locus of the RGB is bluer than that of globular clusters with the same
RGB slope suggesting that field RGB stars are about $3$ Gyr old.

Encouraged by our earlier near--infrared study of the Magellanic
Clouds (Cioni et al. \cite{tip}, \cite{morf}, Cioni \& Habing
\cite{cm}) we obtained comparable observations in the $IJK_{\mathrm
  s}$ bands of the central $20^\prime \times 20^\prime$ region of NGC
6822. Our survey, slightly less extended than that of Letarte et al.
(\cite{tarte}), quadruples the number of AGB stars observed by Gallart
et al. (\cite{galla1}). Strategy and details of the
observations are reviewed in Sect. 2 while results are presented in
Sect. 3. In Sect. 4 the distance, the metallicity and other properties
are discussed and compared with the existing literature. Sect. 5
concludes our study. This paper is the first from a project devoted to
late--type stars of northern Local Group galaxies started by Habing \&
Cioni in 2002. The results obtained from other targeted galaxies (less
populated in AGB content) will be presented in a subsequent paper.

\section{Observations}

Observations were performed with the 4.2m William Herschel Telescope
(WHT) on La Palma (Spain) during two runs in July 2002. The Isaac
Newton Group Red Imaging Device (INGRID) was used to observe in the
$J$ and $K_{\mathrm s}$ band from 20th to 22nd July while observations
in the $I$ band were obtained using the Prime Focus imaging camera
(PFIP) on July 14th. INGRID is a near--infrared (near--IR) camera for
use at the Cassegrain Focus. It has a $1024\times 1024$ Hawaii
near--IR detector with a pixel scale of $0.238$\arcsec /pix which
gives a total field of view of $4.06$\arcmin $\times 4.06$\arcmin. The
PFIP is an optical mosaic camera of two EEV $2$k$\times 2$k CCDs
giving a pixel scale of $0.24$\arcsec /pix and a field of view of
$16.2$\arcmin $\times 16.2$\arcmin. A central gap between the two
chips leaves a gap of $9$\arcsec ~between the two I--band images.
Observations in the three wave bands have been taken as close as
possible in time so that the colours will not be affected by the
variability of the sources. The exposure time in $I$ was $900$s while
$J$ and $K_{\mathrm s}$ images were obtained exposing for $2$s each of
$4$ co-averaged images of a single dithered position out of $5$.

The total integration time allowed us to reach S/N$>3$ in photometric
measurements as faint as about $1$ mag below the tip of the red giant
branch (TRGB) and to detect stars as bright as $2$ mag above the TRGB
without detector saturation. In total we covered an area of about
$20^\prime \times 20^\prime$ centered at $\alpha = 19$:$44$:$56.6$ and
$\delta = -14$:$47$:$21$. On average three photometric standard star
fields were observed each night in each wave band. Near--IR
photometric standards were taken from the ARNICA list (Hunt et al.
\cite{hunt}) and $I$--band photometric standards from Landolt
(\cite{land}).  The sky was clear during both runs; the FWHM of the
stars indicates that the seeing was below $1^{\prime\prime}$ in the IR
wave bands and about $1.3^{\prime\prime}$ in the $I$ band.  In the
near--IR sky frames were observed after each pointing on NGC 6822 in
directions about $20^{\prime}$ away in Right Ascension.

During the observations we noticed  that the upper right corner of the
INGRID field  presented non--negligible distortions.  Subsequently the
overlap among the images of the mosaic was increased to cover at least
a quarter of the field of view.

\section{Data reduction}
The processing of raw images was done using the IRAF software.

\subsection{$J$ \& $K_{\mathrm s}$ bands}

We used the INGRID quick look package available at
(http://\-www.\-ast.\-cam.\-ac.\-uk\-/ING/\-Astronomy/\-instruments
\-/ingrid/\-ingrid\_ql.\-html) to split pre--read and post--read
images and to apply the flat--field correction and the sky subtraction
({\it iframediff}).  The flat field frame in each wave band was
created from twilight exposures and the subtracted sky frame was
flatfielded before removal. Dithered images were then aligned and
combined. Dark current correction is implicitly performed during the
sky subtraction.

A mosaic of all images in one wave band was created by matching
overlapping regions between the fields.  This process reduced
considerably the effect of image distortions with array position. The
two mosaics were finally aligned with each other.

\subsection{$I$ band}

The standard bias and flatfield correction was applied to each chip.
Fringes were removed by dividing each time--normalized resulting image
by a fringe time--normalized image. The latter (star free) image was
created by combination of six images centered on the sparse Local
Group galaxy Draco; for each image we used an exposure time of $300$s
and we corrected for bias and flat field. The observations of the
Draco fields were obtained close in time ($900$s on average) to the
observation of NGC 6822.

\subsection{Astrometry}
Accurate positions have been derived in each wave band using a few
stars from the Guide Star Catalogue II (GSC2.2,
http://www-gsss.stsci.edu/gsc/gsc2/GSC2home.htm). Astrometric
information was present in the fits  header of each near--IR
mosaic image. Using Skycat (http://archive.eso.org/skycat/) we
first identified a few stars  from  GSC2.2 and then  used IDL
routines (STARAST) to correct  their astrometry. $I$--band  images
did not have the proper keywords in their fits header  and these
keys were reconstructed in a similar way using the same GSC2.2
reference stars.   At last near--IR mosaic  images were compared
to  the $I$--band images to obtain a proper overlap between
sources prior to their extraction.

\begin{figure}
\resizebox{\hsize}{!}{\includegraphics{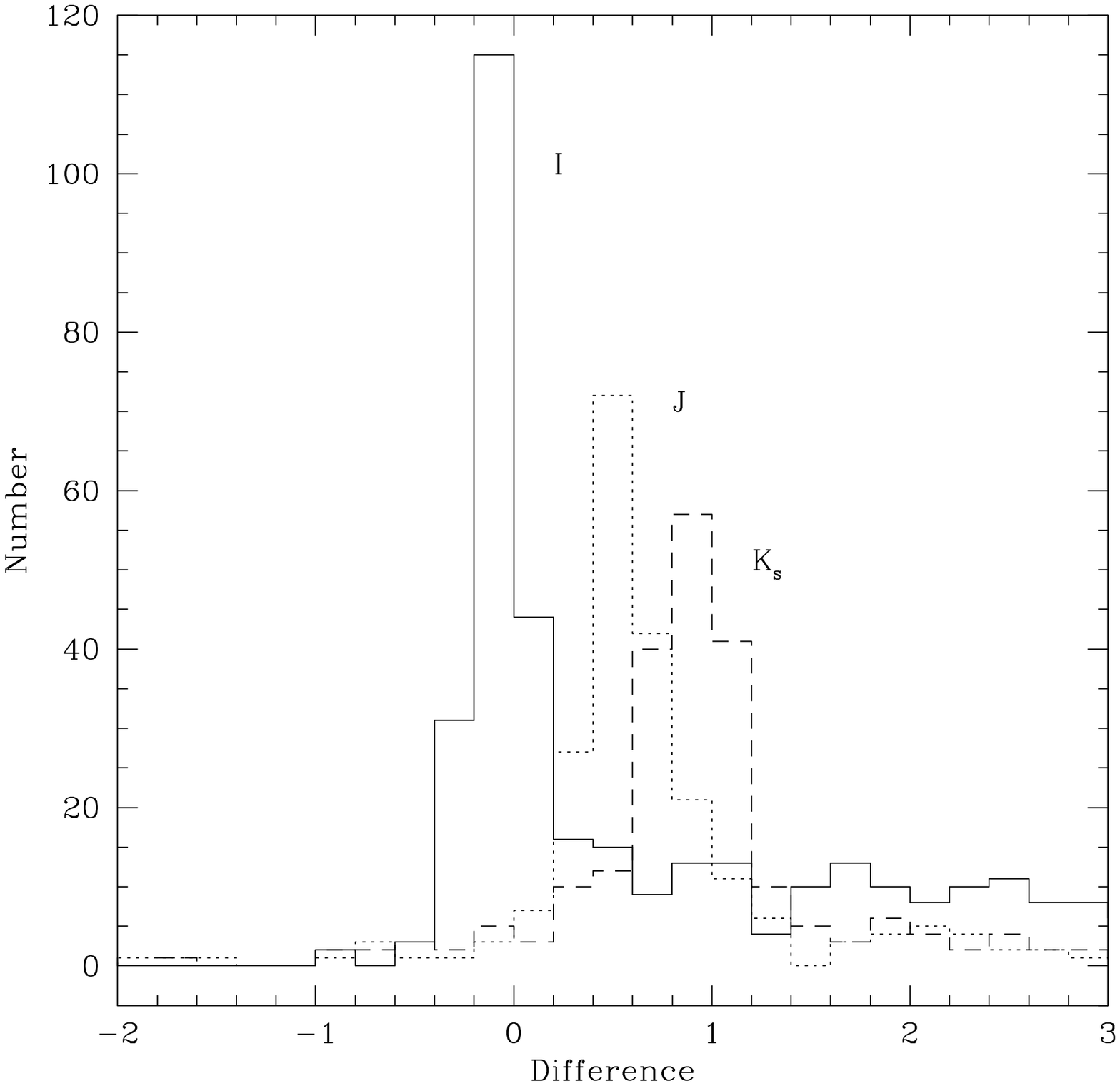}}

\caption{Magnitude difference between the $J$ (dotted histogram) and 
  $K_{\mathrm s}$ (dashed histogram) photometry derived in this paper
  and the 2MASS photometry of $461$ objects, and $557$ DENIS stars in
  the $I$ band (continuous histogram). Note the tail at large
  differences that indicates the presence of variable stars. Binning
  is $0.2$ mag.}

\label{shift}
\end{figure}

\subsection{Photometry}

Sources were extracted using the SExtractor program (Bertin \& Arnout
\cite{sex}).  The flux of each source was calculated for an aperture
of $5$ pixels and for a detection and analysis threshold equal to
$1.5$; except for the parameters that characterize the detector we used
default values for the remaining keywords in the SExtractor
configuration file.

The source extraction was performed first in each wave band
separately. Afterwards we matched the $J$ and $K_{\mathrm s}$
detections using the full mosaic images and an association radius of
$4$ pixels. The nearest source was kept as a counterpart.  The
cross--identification of $I$--band sources was made separately for the 
$I$--band chip using the near-IR registered images obtained from the
$J$ and $K_{\mathrm s}$ mosaics. Sources were matched using the same
criteria. The combination of seeing and crowding in the $I$ band did
not allow us to resolve accurately the stars towards the very center
of the galaxy; we have missed some as bright as AGB stars. These
sources, however, have been detected in the $J$ and $K_{\mathrm s}$
wave bands.

SExtractor assigns to each extracted object a flag that depends on the
quality of the photometric detection.  Flag values are given in the
documentation available at
(http://terapix.iap.fr/rubrique.php?id\_rubrique=91).

The photometric calibration was performed for each observing night.
We also used SExtractor using all standard stars present in each
field. Aperture of $20$ and $15$ pix were used to collect most of
the flux in the $I$ and $JK_{\mathrm s}$ bands, respectively. Multiple
measurements were averaged prior to the exclusion of a few outliers (i.e.
sources with too few counts). Constant extinction coefficients in the
near--infrared bands of $0.1$ mag in $J$ and $0.12$ mag in $K_s$ were
taken from the intrument web page.  The resulting zero--points are
$I=23.30\pm0.09$, $J=23.02\pm0.10$ and $K_s=22.48\pm0.13$. The
near--IR zero--points are about $0.4$ mag fainter than those indicated
in the instrument web page, therefore observations discussed in this
paper did not take place during perfect photometric conditions.

In order to improve the calibration of our data we have
cross--identified them with the 2MASS
(http://\-www.\-ipac.\-caltech.\-edu\-/2mass/) and DENIS catalogue
(http://\-cdsweb.\-u-strasbg.\-fr/\-denis.\-html).  We restricted the
cross-identification to those sources covered by one of the $I$--band
chips. Because the great majority of our targets is variable (c.q. AGB
stars) we considered, among sources detected in three wave bands, only
those with: $K_{\mathrm s}<17$ and $J-K_{\mathrm s}<0.8$.  These
should be mostly foreground objects but a few supergiants, bright AGB
stars and other young stars will have been included. Coordinates in
all three catalogues agreed very well with each other, providing only
a single counterpart within $3^{\prime}$; most of the sources differ
by not more than $40^{\prime \prime}$. We obtained $461$ objects in
common with the 2MASS catalogue and $557$ objects in common with the
DENIS catalogue of which $230$ and $426$ have a distance below
$40^{\prime \prime}$. The latter were used to calculate the systematic
photometric differences between the catalogues (Fig. \ref{shift}).
Resulting values are:

$I = I^{DENIS} - 0.1$

$J = J^{2MASS} + 0.5$

$K_s = K_{\mathrm s}^{2MASS} + 0.9$ 

These shifts have been applied to all extracted sources in NGC 6822.

\begin{figure}
\vspace{9cm}
\caption{Photometric errors in $I$ (a), $J$ (b) and $K_{\mathrm s}$ 
  (c) versus magnitude. The two curves in (a) refer to chip 1 (upper)
  and 2 (lower) $I$--band observations. Matched $J$ and $K_{\mathrm
    s}$ sources have photometric errors as indicated by the two higher
  curves in (b) and (c) while the third, lower, curve indicates single
  band detections in $J$ and $K_{\mathrm s}$, respectively. 
  Vertical lines discriminates between all extracted sources and 
  sources included in the final catalogue.}

\label{err}
\end{figure}

\subsection{Catalogue}

Table \ref{cat} shows the first ten lines of the full table ({\it
  table.dat}) that is only available in the electronic edition of this
article. The table contains $9377$ stars detected in all three bands,
$I, J$ and $K_{\mathrm s}$: columns $1$ and $2$ list Right Ascension
and Declination in degrees at epoch J2000, columns $3$, $4$ and
$5$ list $I$ magnitude, photometric error and SExtractor flag,
respectively; columns $6-8$ and $9-11$ contain the same information
for the $J$ and $K_s$ wave bands, respectively; this table lacks sources
in the gap between the two $I$--band fields. A second table ({\it
  table2.dat}) contains similar information for $16354$ sources
detected in both the $J$ and $K_{\mathrm s}$ bands but not in the
$I$-band, and in the full mosaic without the central gap imposed by
the $I$--band data. Note that some sources from {\it table.dat}
also occur in table {\it table2.dat}. However we decided to release both
tables separately for the following reasons: after aligning the
near--IR images to the area(s) covered by the $I$--band 
the resulting photometric
errors appear slightly different (cf. Fig.2), and the astrometric solution
is more precise in the three--band catalogue.  {\it Table2.dat} is
released to support further studies that do not require a match with
the $I$--band data but instead reliable photometric correspondence
between the two near--IR bands that also includes the very central
region of the galaxy.

In both tables only sources with SExtractor flag $<4$ in at least two
wave bands have been included, with the additional condition that
$J<20$ and $K_{\mathrm s}<18$ (this implies $I<22.5$) and to exclude
saturated objects: $I>16.5$.  This represents a suitable compromise
between reliability and completeness.  Fig. \ref{err} shows the
behaviour of photometric errors for all target stars as a function of
magnitude in each wave band and indicates the region of released data.
Sources detected only in one wave band or sources that were
excluded by the release selection criteria will be made available upon
request to the first author.

\begin{table*}
\caption{NGC 6822 catalogue of sources detected in three wave bands.}
\label{cat}
\[
\begin{array}{ccccccccccc}
\hline
\noalign{\smallskip}
\alpha & \delta & I & \mathrm{e}(I) & \mathrm{f}(I) & J &  \mathrm{e}(J)  & \mathrm{f}(J)  &  K_s  & \mathrm{e}(K_s)  &
\mathrm{f}(K_s) \\
296.104218 & -14.921373 & 20.32 & 0.38 &  2 & 19.25 & 0.42 &  2 & 17.81 & 0.23 &  2 \\ 
296.217651 & -14.921238 & 19.33 & 0.25 &  2 & 18.37 & 0.32 &  2 & 17.63 & 0.23 &  0 \\ 
296.190765 & -14.920881 & 21.20 & 0.53 &  0 & 19.70 & 0.47 &  2 & 17.94 & 0.24 &  0 \\ 
296.103973 & -14.921098 & 20.32 & 0.38 &  2 & 19.25 & 0.42 &  2 & 17.93 & 0.24 &  2 \\ 
296.107574 & -14.919714 & 20.21 & 0.36 &  0 & 18.84 & 0.37 &  0 & 17.93 & 0.24 &  2 \\ 
296.210419 & -14.919394 & 19.29 & 0.24 &  2 & 18.24 & 0.31 &  2 & 17.45 & 0.22 &  0 \\ 
296.164001 & -14.920863 & 19.04 & 0.22 &  0 & 17.45 & 0.23 &  0 & 16.83 & 0.20 &  0 \\ 
296.188293 & -14.920278 & 19.53 & 0.27 &  0 & 17.34 & 0.22 &  0 & 16.02 & 0.16 &  0 \\ 
296.165131 & -14.919350 & 18.79 & 0.19 &  2 & 17.67 & 0.25 &  0 & 17.31 & 0.22 &  0 \\ 
296.179962 & -14.919452 & 18.61 & 0.18 &  2 & 17.29 & 0.21 &  0 & 16.69 & 0.19 &  0 \\ 
\noalign{\smallskip} \hline
\end{array}
\]
\end{table*}

In our near--IR observations the confusion by merging stellar images
is substantially smaller than in maps of comparable angular
resolution but taken in other spectral ranges. The contamination by 
unresolved background sources and by sources of a different nature
or age, is strongly reduced and the completeness of $J$ and $K_s$ band
images is almost 100\%; at the faintest magnitudes the completeness is
dominated by seeing variations. This follows from inspection by eye
after plotting the extracted sources on to the images. Sources with
magnitude around or brighter than the TRGB are not affected by
confusion. All this is not true for our $I$--band images that are
incomplete in a central field of about $5^\prime \times 5^\prime$, due
to the effects of both crowding and seeing (note that the observations
were obtained with a seeing of about $1.3^{\prime\prime}$). The
incompleteness of the $I$--band detections can be estimated using
sources detected in the $J$ and $K_s$ wave bands only. We find $95$\%
completeness at $I=19.0$, $93$\% at $I=19.5$, $80$\% at $I=20.0$ and
below $50$\% for $I>20.5$.  Up to about $I=19.5$ we detected $91$\%
and $98$\% of the sources seen in the $J$ and $K_{\mathrm s}$ bands.
We attribute the missing $9\%$ and $2\%$ to crowding. At magnitude
$I>19.5$ the completeness begins to decrease monotonically, which is
probably due to missing faint sources in the $J$ and $K_{\mathrm s}$
bands.

\section{Results}
\subsection{Magnitudes and Colours}

\begin{figure}
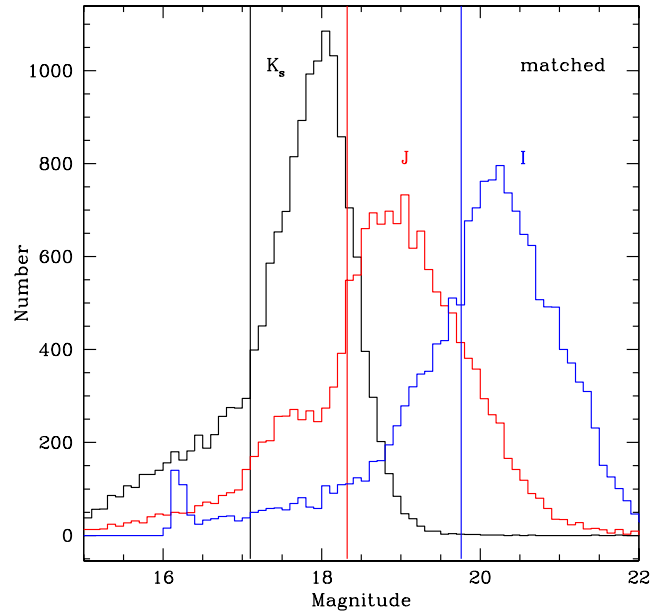

\vspace{9cm}
\caption{Combined $J$ and $K_{\mathrm s}$ of the central $22^{\prime}$ 
  square of NGC 6822.  North is up and East to the left. This
  image is centered at $\alpha = 19$:$44$:$56.6$ and $\delta =
  -14$:$47$:$21$. The colour version of this image can be seen in the
  electronic edition of this article.  This colour version shows a few
  very red stellar objects detected only in $K_{\mathrm s}$. These are
  candidate AGB stars with thick circumstellar envelopes.}
\label{image3}
\end{figure}

\subsubsection{$I$, $J$ and $K_{\mathrm s}$ histograms}
\begin{figure}
\resizebox{\hsize}{!}{\includegraphics{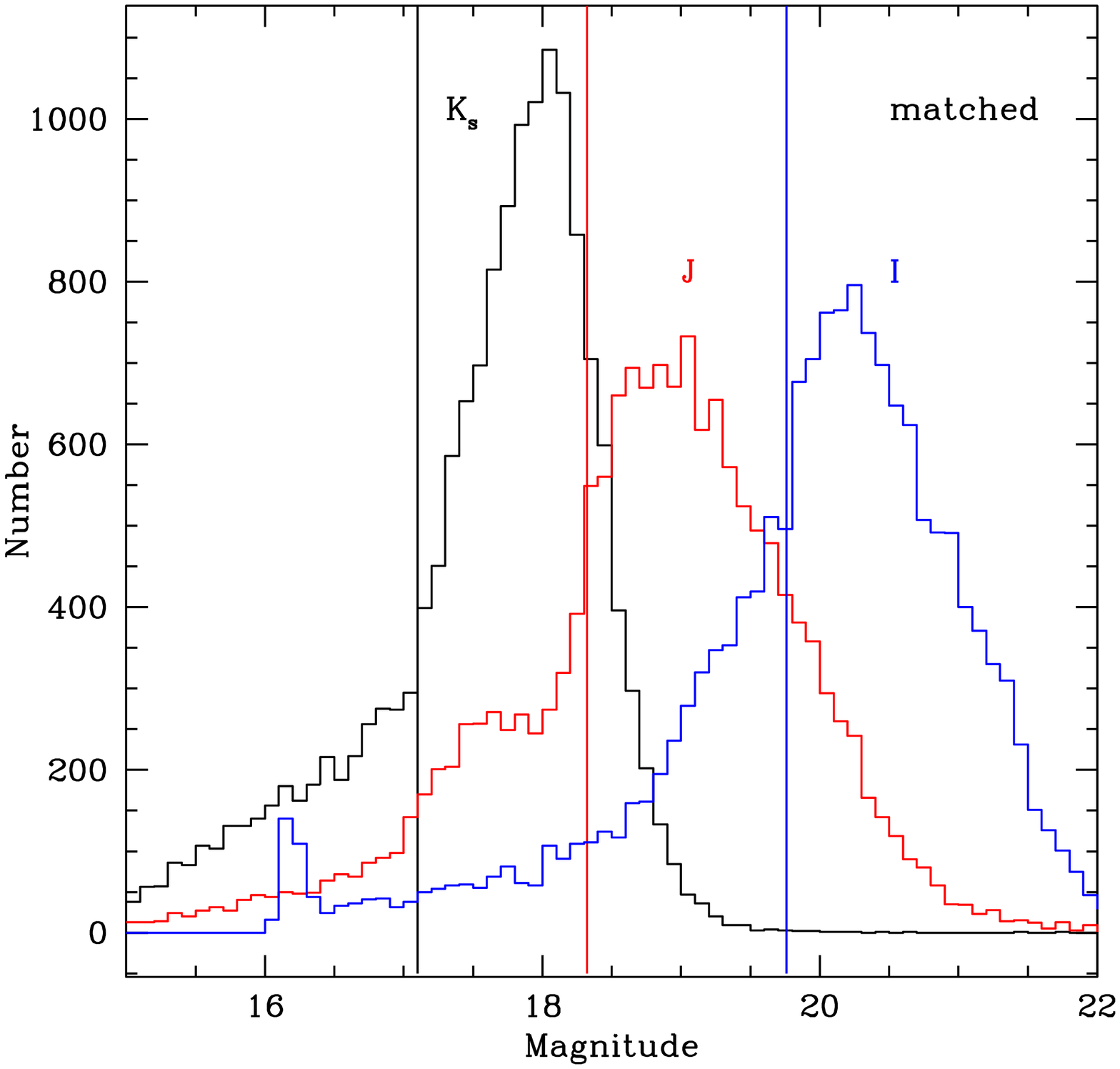}}
\caption{Histogram of the number of sources matched among the three
wave bands as a function of magnitude. Vertical lines show the
location of the TRGB. Binning is $0.1$ mag. The little bump at about 
$I=16$ is caused by $I$--band saturated objects that have been excluded 
from the final catalogue.} 
\label{hist}
\end{figure}

Figure \ref{hist} shows the histogram of the number of sources
detected in all three wave bands as a function of magnitude. Vertical
lines indicate the position of the TRGB.  Although the discontinuity
produced by the TRGB is clearly visible we determined a more precise
location with the same technique as used in Cioni et al. (\cite{tip}).
The TRGB is at $I=19.76\pm0.01$, $J=18.32\pm0.01$ and $K_{\mathrm
  s}=17.10\pm0.01$. Errors equal $\sigma_M/\sqrt{N}$ in each band
where $\sigma_M$ is the mean of all errors within $0.1$ mag from the
TRGB location and $N$ is the total number of sources contributing to
$\sigma_M$.  Sources brighter than the TRGB are mostly AGB stars while
fainter sources are RGB stars mixed with early--AGB (EAGB) stars; the
RGB giants outnumber the EAGB stars by so much that we will
ignore the latter category. We did not subtract the foreground stars
from the histograms prior to the determination of the TRGB position
because their number is small enough to be ignored: the foreground
contribution in the $I$ band has been checked by assuming that stars
located in a small box on each edge of the total survey area are all
foreground stars and in the $J$ and $K_{\mathrm s}$ bands off--set
fields observed for image--sky--subtraction were used to estimate the
foreground contribution. As a test we subtracted the foreground
contribution from the magnitude histograms before the determination of
the TRGB and we obtained the same result. Note that the little bump at
about $I=16$ is caused by objects that saturated the detectors and
that have been excluded from the released catalogue.

\subsubsection{($I-J$, $J-K_{\mathrm s}$) colour--colour diagram}
\begin{figure}
\resizebox{\hsize}{!}{\includegraphics{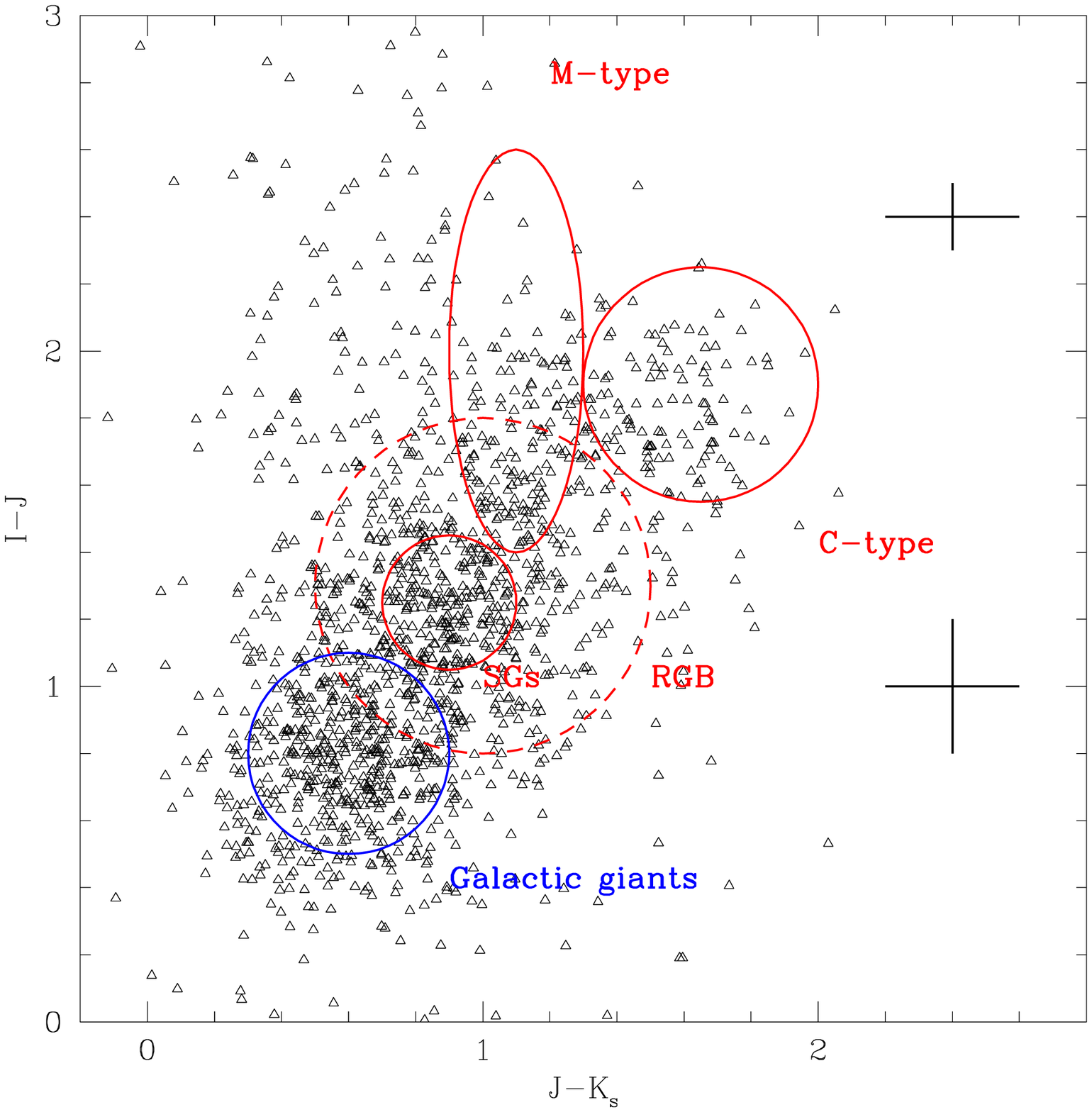}}
\caption{Colour--colour diagram of sources detected in NGC 6822 in three wave
  bands and with photometric errors below $0.22$ mag in each band.
  Four regions that contain statistically different types of sources
  are indicated as well as the location of galactic giant stars. Crosses 
  show the maximum errors in the two colours.}
\label{col}
\end{figure}

The colour--colour diagram (Fig. \ref{col}) contains only sources with
a photometric error below $0.22$ mag in all three wave bands.  These
stars are representative of the distribution of sources in each region of the
diagram; there is a negligible contamination by faint RGB stars.
Sources with $I-J<1.0$ and $J-K_{\mathrm s}<0.8$ are mostly foreground
sources.  Although the low latitude of NGC 6822 suggests that there is
 considerable foreground contribution, foreground stars have much
lower $J-K_{\mathrm s}$ colours and hence they do not overlap with the
distribution of the AGB population.  The middle dashed circle contains
practically all RGB stars belonging to NGC 6822. Our experience with
the red giants in the LMC and SMC shows that stars with $I-J\approx
1.2$ and $J-K_{\mathrm s}\approx 1.0$ have early M--type spectra and
thus an atmosphere with a C abundance less than the O abundance.  Those
with later M--type spectra extend to larger $I-J$ colours but at
approximately the same $J-K_{\mathrm s}$ colour. Stars with
$I-J\approx 1.8$ and $J-K_{\mathrm s}\approx 1.5$ are C--rich AGB
stars. Those with $J-K_s>2$ are candidate obscured AGB stars and they
can be either O--rich or C--rich. These locations depend somewhat on
extinction and metallicity and are discussed here as indicative of
which type of object is found around a given location in the
colour--colour plot.  Note that photometric errors blur the
distribution of sources in the various regions.  In fact it is
difficult to identify red supergiants that are probably present in
this galaxy (middle continuous circle).  For a more detailed
discussion and visualization of the regions in the colour-colour
diagram see Cioni et al. (\cite{mess}).

\subsubsection{($I-J$, $I$) colour--magnitude diagram}
The colour--magnitude diagram ($I-J$, $I$) shown in Fig. \ref{iji} is
a suitable tool for discriminating AGB stars from other stars in the
galaxy and in the foreground (Cioni et al. \cite{morf}).  The
horizontal line marks the position of the TRGB while the slanted line
($I=-4.74\times(I-J)+24.15$) distinguishes between RGB plus AGB stars
and either foreground or younger stars of NGC 6822. The slope of the
slanted line is the same as in Cioni et al.  (\cite{morf}). Its
zero-point has been adjusted by eye to be tangential to the plume of
AGB stars. A few tests were made to select a zero-point that would
compromise between including red-supergiants and excluding AGB stars
of early spectral sub--type.  Objects with $I-J=1.6$ and $I=17.6$ are
probably red--supergiants.  The largest concentration of objects
is that of the upper RGB stars while AGB stars populate the plume of
objects above the TRGB. For increasing envelope thickness the AGB
plume bends down to faint magnitudes and red colours and obscured AGB 
stars may have an $I$--magnitude below that of the TRGB.

\begin{figure}
\vspace{9cm}
\caption{Colour--magnitude diagram of sources detected in all
  three wave bands. Two lines separate the stars into three regions. AGB
  stars are above the TRGB and redder than the slanted line
  $I=-4.74\times(I-J)+24.15$; RGB stars have similar colours as AGB
  stars but they are fainter and below the TRGB. Genuine, younger NGC
  6822 stars are found on the blue side of the slanted line together
  with foreground stars.}
\label{iji}
\end{figure}

\subsubsection{($J-K_{\mathrm s}$, $K_{\mathrm s}$) colour--magnitude diagram}
AGB stars selected as above are plotted in the colour--magnitude
diagram ($J-K_{\mathrm s}$, $K_{\mathrm s}$) in Fig. \ref {jkk}. The
distribution shows a rather good correlation between colour and
magnitude: when the colour increases the magnitude increases as well,
but the correlation is apparent and not real. Analysis of similar data
in the LMC and SMC shows that the diagram contains two sequences: one
vertical sequence around $J-K_{\mathrm s}= 1.0$ that consists of stars
with M--type spectra (O--rich stars) and a separate branch that
departs to redder colours from the vertical O--rich branch around
$K_{\mathrm s}= 17$ to $16$; this second branch contains C--rich AGB
stars. The histogram in the upper right--hand corner indicates more
precisely the separation between the two branches. It is the
histogram of $J-K_{\mathrm s}$ colours for sources having $K_{\mathrm
  s}<16.5$.  The peak of the distribution corresponds to O--rich
stars. On the red side of the peak there is a significant drop and
then a plateau; this plateau contains the C--rich AGB stars.  In the
histogram the vertical line indicates the adopted separation between
the two groups of stars.  The same line is shown in the main part of
the figure. Note that this separation has been applied to the data 
independently of their $K_{\mathrm s}$ magnitude.   
The use of this selection criterion may be a conservative approach in
defining the number of C--rich AGB stars (cft. Raimondo et al.
\cite{gabri}) but intrinsic C--rich AGB stars are usually brighter
than TRGB stars and those C--rich stars below the TRGB are either
extrinsic carbon stars (they belong to a binary system and have been
polluted by AGB winds) or are sources with larger photometric errors
that should instead be on to the O--rich branch.  Therefore we
adopted $J-K_{\mathrm s}=1.36$ as the colour that distinguishes
between sources of different chemical types.  There are $500$ and
$2161$ C--rich and O--rich stars, respectively, a ratio of $0.23$. 

The adopted criterion for distinguishing between C-- and O--rich stars
implies a sharp transition between the two regimes. In a composite stellar 
population variations both in age and metallicity throughout the galaxy 
are expected. The $J-K_{\mathrm s}$ colour of the giant branch varies 
by about $0.3$ mag for a variation in [Fe/H] of about $1$ dex 
(Ivanov \& Borissova \cite{ivi}). A variation of about $0.3$ mag is also 
predicted in the $K_{\mathrm s}$ band for ages above $2$ Gyr, the variation 
in the $J$ band is somewhat less, inducing a variation in 
the $J-K_{\mathrm s}$ 
colour of about $0.1$ mag (Girardi et al. \cite{leo}). Differential 
reddening will also shift the giant branch to red colours. Moreover most 
AGB stars above the TRGB are long-period variables and vary with an amplitude 
of at least $0.1$ mag in the $K_{\mathrm s}$ band (Cioni et al. \cite{iso}).
We conclude that the combined effect of age, 
metallicity  but also of extinction and variability 
in the $J-K_{\mathrm s}$ colour is of the order of the photometric 
errors involved in this study. In Sect. 4.2.3 we show the effect of 
changing the selection criterion on the determination of the C/M ratio.

\begin{figure}
\vspace{9cm}
\caption{Colour--magnitude diagram of AGB stars detected in three
wave bands. Vertical lines indicate the separation between O--rich
and C--rich stars. The histogram of the number of AGB stars versus 
$J-K_{\mathrm s}$ colour for $K_\mathrm s<16.5$ is shown in the upper 
right corner. The plateau at about $J-K_s=1.5$ shows the contribution 
of C--rich stars. Binning is $0.1$ mag. 
}\label{jkk}
\end{figure}

Figure \ref {jkk} contains O--rich AGB stars at magnitudes well below
the $K_{\mathrm s}$ magnitude of the TRGB. We know that they are AGB
stars because of their $I-J$ colour; they would not have been
recognized as such had we had only $J$ and $K_{\mathrm s}$ magnitudes:
The $K_{\mathrm s}$ magnitude of the TRGB depends on colour and is
horizontal only in the $I$--band. On the other hand, selecting AGB
stars only in the ($I-J$, $I$) diagram excludes most obscured AGB star
candidates, because they would have an $I$--band magnitude below the 
TRGB, for which a spectral type is rather uncertain and needs
spectroscopic mesurements (up to about 50\% of these red stars could
be O--rich).

\subsubsection{$(I-J)_0$ histogram}
\begin{figure}
\resizebox{\hsize}{!}{\includegraphics{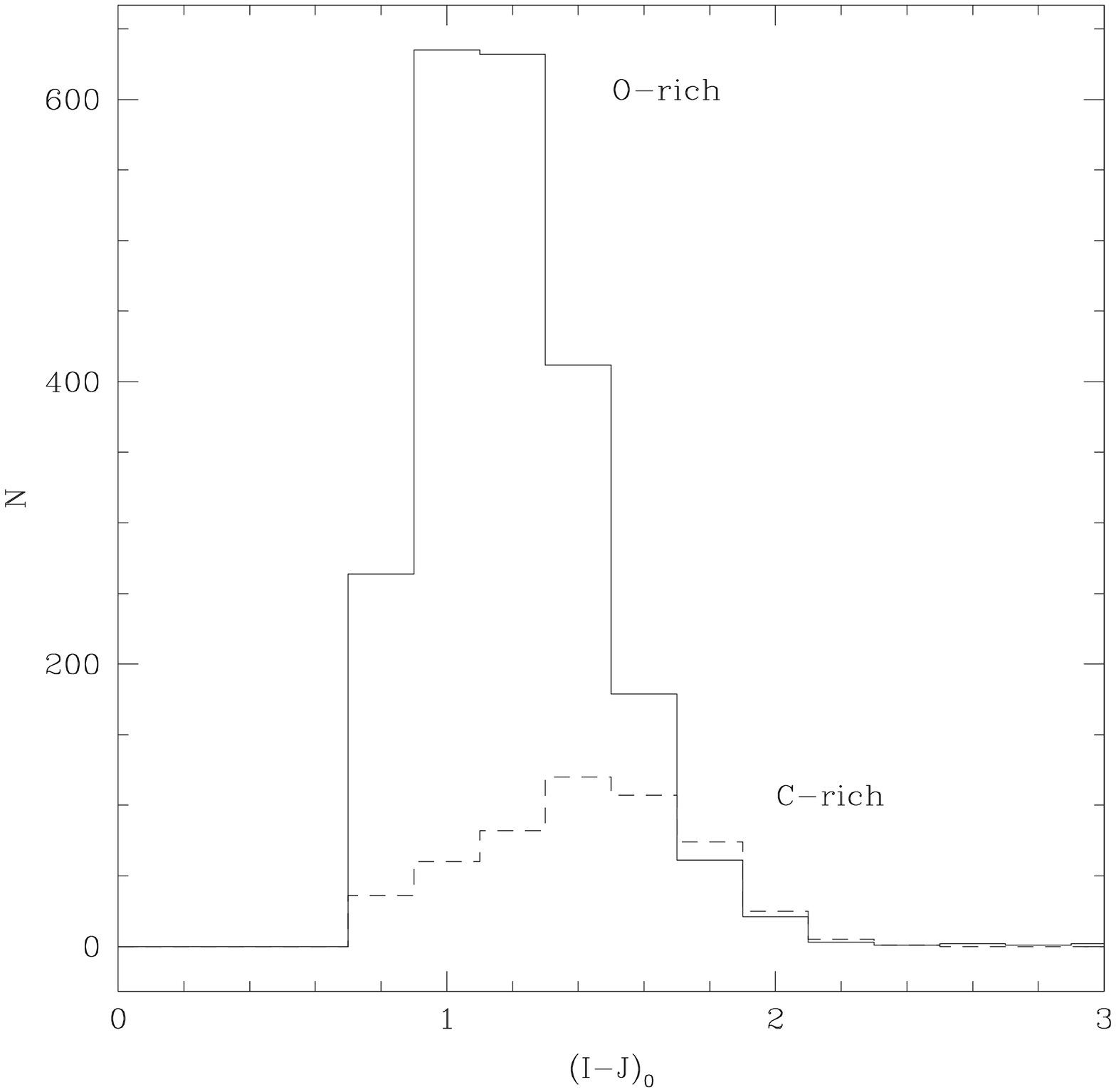}}
\caption{Histogram of $(I-J)_0$ colour of O--rich (continuous line)
  and C--rich (dashed line) AGB stars detected in three wave bands in
  NGC 6822. This colour is a function of the M spectral subtype of
  O--rich stars. Bins have a width of $0.2$ mag.}
\label{sp}
\end{figure}

The $(I-J)_0$ colour of the AGB stars is plotted in Fig. \ref{sp}; the
reddening correction is discussed in Sect. 5.1.  This histogram shows
the distribution, as a function of colour, of O--rich and C--rich AGB
stars, respectively.  For O--rich stars the variation in $(I-J)_0$
colour is correlated with the various sub--type M--spectra (M0, M1, M2
etc.; see Glass \& Schultheis \cite{glsc}, Fluks et al \cite{fluks},
Blanco et al. \cite{blanco}, Blanco \& McCarthy \cite{blca}): this
correlation is caused by the TiO molecular absorption bands that
dominate the photometric $I$--band: the TiO bands increase in
strength considerably when the atmosphere gets cooler. Compared to the
Magellanic Clouds (Cioni \& Habing \cite{cm}) most O--rich stars in
NGC 6822 have M spectral sub--types between M0 ($(I-J)_0\approx 1$)
and M4 ($(I-J)_0\approx 1.4$). C--rich stars follow a rather symmetric
distribution peaking at about $(I-J)_0=1.4$.

\subsection{Spatial distribution}
\subsubsection{Sources matched among three wave bands}
\begin{figure*}
\centerline{%
\epsfxsize=0.5\hsize \epsfbox{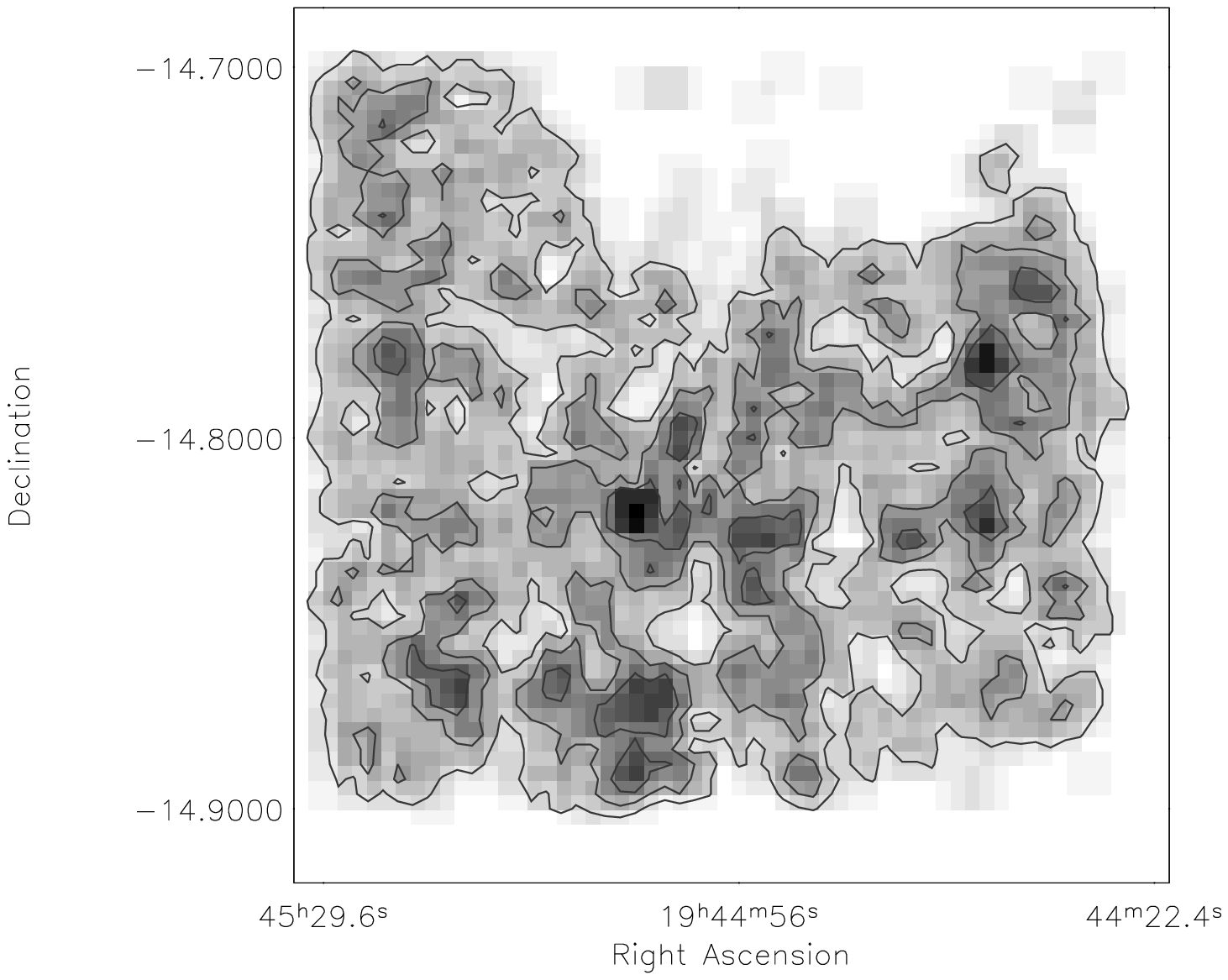}\quad
\epsfxsize=0.5\hsize \epsfbox{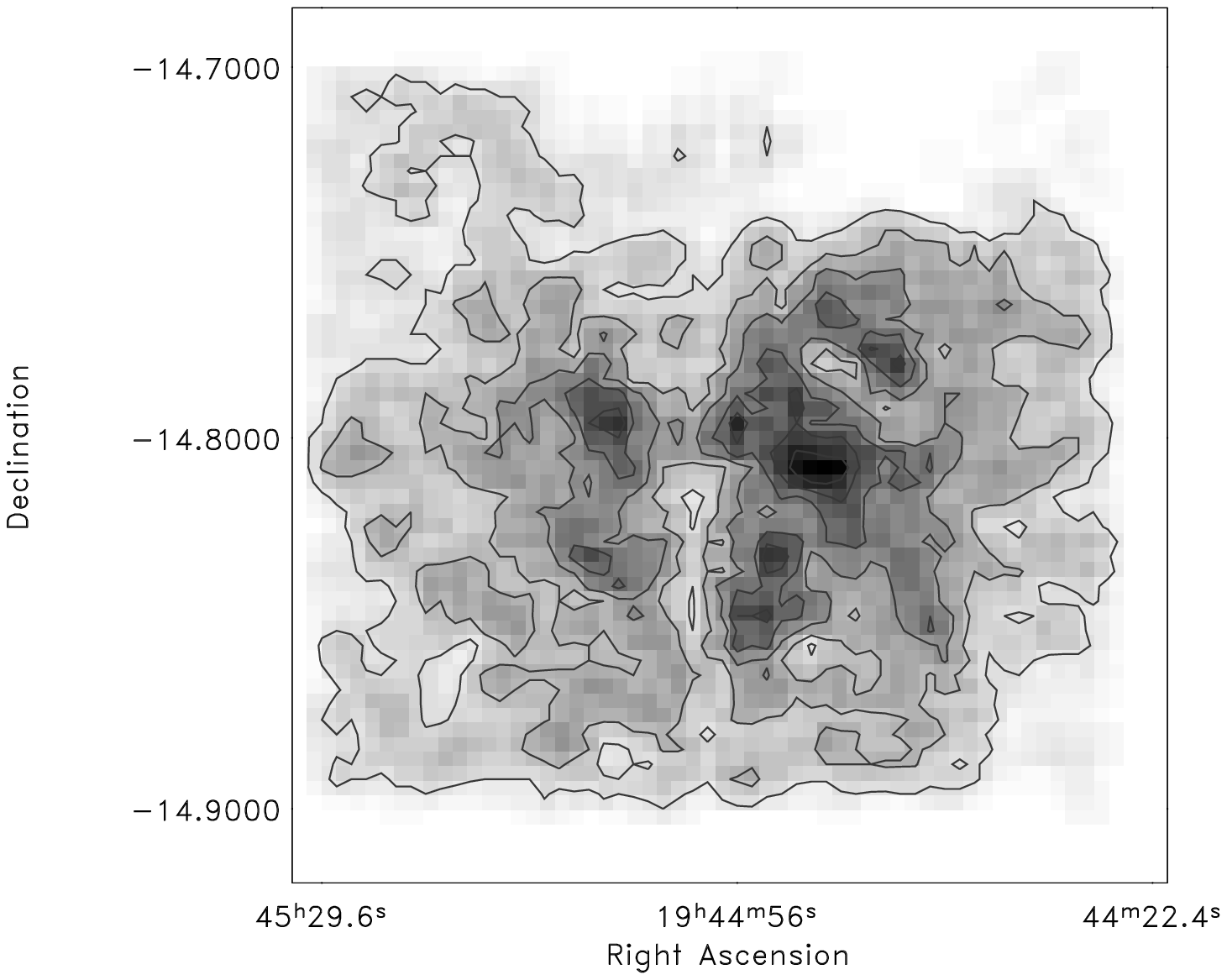}}
\smallskip
\centerline{%
\epsfxsize=0.5\hsize \epsfbox{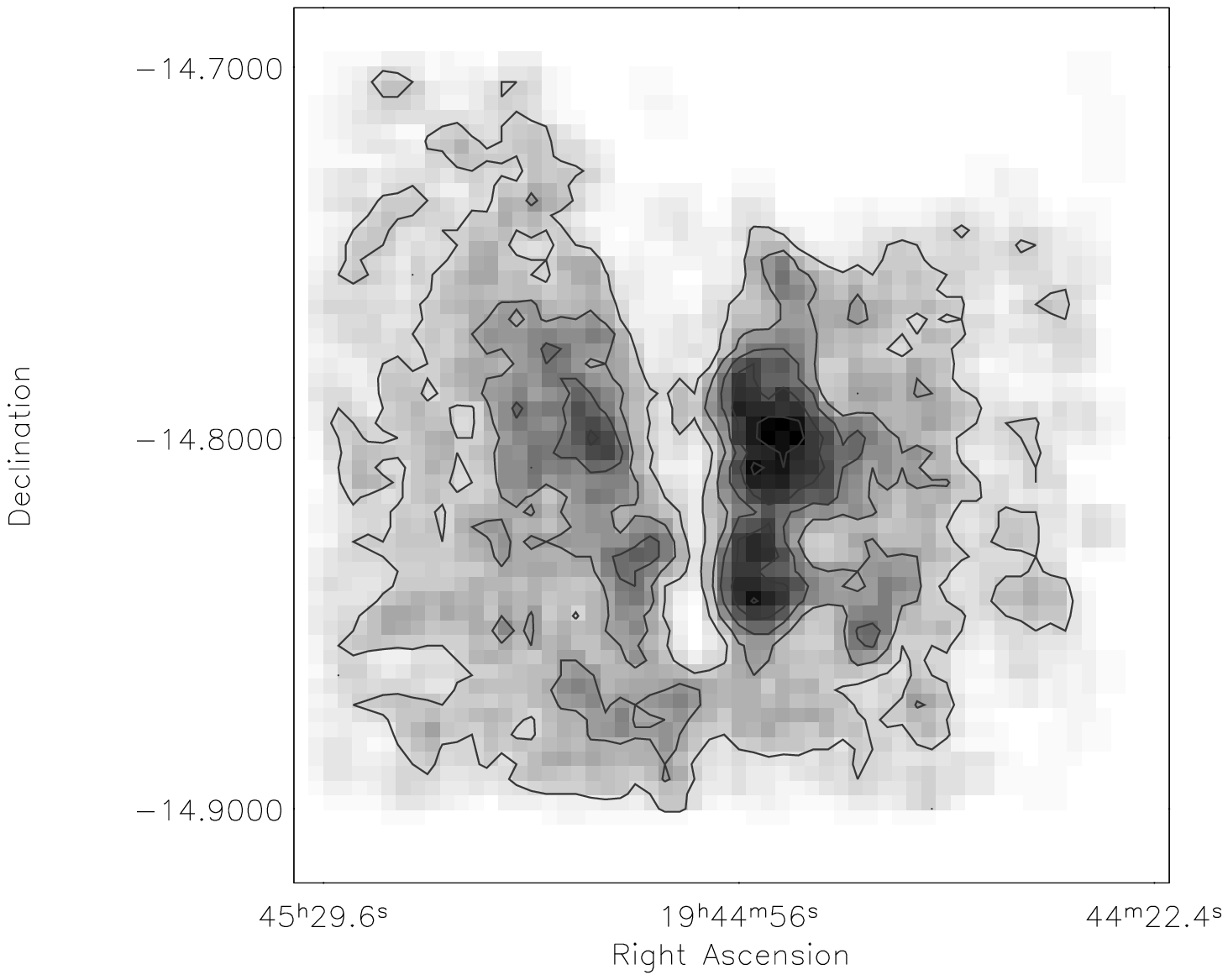}\quad
\epsfxsize=0.5\hsize \epsfbox{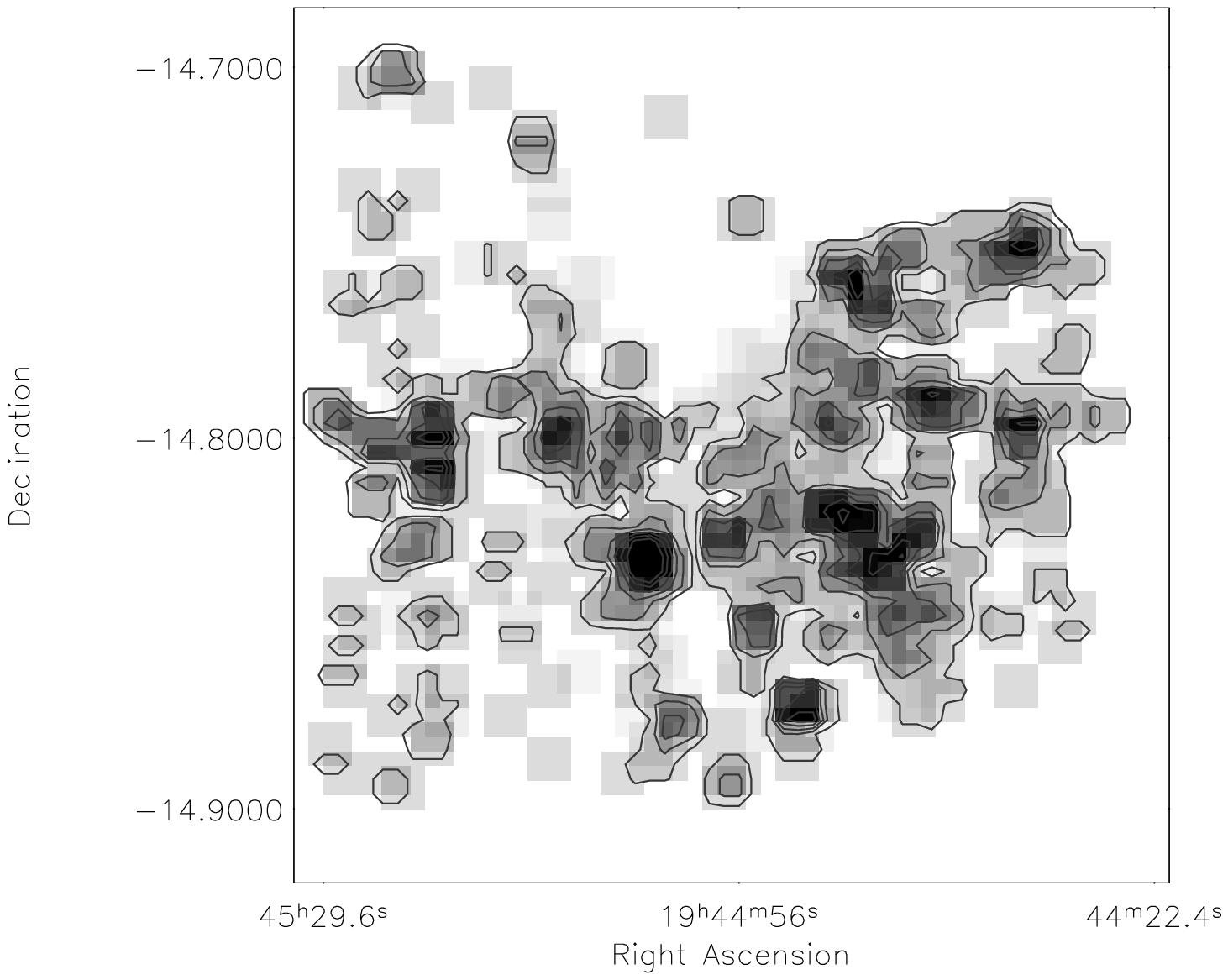}} 
\caption{Logarithmic and smoothed density distributions of stars detected
  in all three bands. Clockwise from top left, we show foreground and young 
  stars, RGB stars, AGB stars and C/M ratio in $60\times 60$ bins
  of about $2^{\prime}$ square each.  
  These maps contain only sources seen in the three
  photometric bands. Darker regions correspond to
  higher density. Contours are: from $2$ to $12$ in steps of $=1$ for RGB
  stars, from $1$ to $9$ in steps of $=1$ for AGB stars, $0.5$, $1.0$,
  $1.5$ for foreground and younger stars and $0.3$,$0.6$,$0.9$ for the
  C/M ratio distribution.  North is up and East is
  left.}
\label{surf}
\end{figure*}

Figure \ref{surf} shows the number density of sources in {\it
  table.dat} (i.e. detected in three wave bands); we have combined
the two chips even though, as discussed above, there is a gap in sky
coverage of $9^{\prime\prime}$. A gap clearly shows-up in Fig.
\ref{surf}. There are, however, also other effects with a bad
influence on the quality of the maps: confusion caused by crowding in
the central area made worse by poor seeing conditions during the
observations.

We counted stars of a given type in $60\times 60$ bins where a single element 
corresponds to about  $2^{\prime}$ square. The source density in each
field has been smoothed using a box car function of width $=2$  
prior to the construction of the grey scale images
where higher concentrations of  sources are indicated by darker
regions. From left to right we show the distribution foreground plus 
younger sources and RGB stars in the first row,  
AGB stars and the distribution of
the C/M ratio, where the number of C--rich and O--rich sources has
been selected as discussed above, in the second row. The lack
of sources at the outer borders is due to a real decrease in the number 
statistics. RGB and AGB stars are 
distributed smoothly and quite regularly around a central, possibly double 
peak, region compared to the more clumply structure described by foreground 
and younger stars. The distribution of the C/M ratio is also rather clumpy.
Contrary to expectations the younger+foreground component does not trace 
the known bar present in this galaxy. Apart from coverage, seeing and 
crowding effects already discussed, it may be that because of the shallow 
limit of the $I$--band observations only a very small young component of 
the galaxy was detected: 
a few upper main--sequence stars and supergiants; the majority being  
below the detection limit. This can also be deduced from Fig. \ref{iji} 
where main--sequence stars, if present,
 would strongly populate the region left of the 
slanted line at $I<20$. At these magnitudes photometric
errors increase and may give rise to a contamination with RGB stars. 
This was also the case in the LMC; compare Figs. 2 and 4  in Cioni \& Habing 
(\cite{morf}).

\subsubsection{Sources matched between $J$ and $K_{\mathrm s}$ only}
If we would have had no $I$--band observations and only $J$ and
$K_{\mathrm s}$ photometry we would obtain the maps in Fig.\ref{surfjk},
which are
similar to those in Fig. \ref{surf}. The area covered and the size of
each bin is the same as in Fig. \ref{surf}, but the bad effects of the
gap, crowding and seeing in the $I$--band data are now of course absent.  
The selection of
RGB and AGB stars has been made in the ($J-K_{\mathrm s}$, $K_{\mathrm
  s}$) diagram. Only stars above the TRGB ($K_{\mathrm s}=17.10$) have
been taken as representative of the AGB population and we
distinguished between O--rich and C--rich using $J-K_s=1.36$ as the
dividing line.  There are $1511$ C--rich stars and $4684$ O--rich
stars; their ratio C/M$=0.32$. The simple
selection criterium ($``(J-K_{\mathrm{s}}>''$ or $``< 1.36''$) does 
not include O--rich AGB stars of early spectral sub--type because 
these are located below the TRGB in the $K_{\mathrm s}$--band and are 
therefore mixed with RGB stars. However, this criterium can be
refined by excluding the bluest and brighter stars (i.e.
red--supergiants and possibly foreground stars) and the reddest stars
(i.e. obscured AGB star candidates of uncertain spectral type) but it
will still produce a biased AGB statistical sample (see Fig.
\ref{cmd}); because of the large number of overlapping sources the
coloured version of this figure is needed to understand the selection 
criteria and is available in the electronic version of
this article. All other sources below the
TRGB were taken to be RGB stars.  Fig.  \ref{cmd} shows also
that the tip of the AGB can be recognized at $K_{\mathrm s}=15$. This
corresponds to $M_K\approx-8.5$ using the distance modulus derived in
Sect. 5.1; this is a rather common upper limit in other galaxies as well.

\begin{figure}
\vspace{9cm}
\caption{Colour-magnitude diagram (available in colour in the electronic 
  version of the paper) of all sources detected in $J$ and $K_{\mathrm
    s}$ regardless of their $I$ band (black); those above the TRGB
  ($K_{\mathrm s}=17.10$) are possibly AGB stars (magenta). AGB stars
  with an $I$--band detection have been selected using two
  criteria: ($I-J$, $I$) and ($J-K_{\mathrm s}$, $K_{\mathrm s}$)
  criteria (blue). Sources with $J-K_{\mathrm s}>1.36$ are likely
  C--rich; the vertical line separates these from 
  O--rich stars.}
\label{cmd}
\end{figure}

Figure \ref{surfjk} shows that the RGB and AGB stars are distributed rather
smoothly over the galaxy. In this figure the most prominent feature
of NGC 6822 is a bar outlined by an increasing number of sources and
by the darkness level. A double-peaked central structure is prominent in the
almost circular RGB distribution, while AGB stars describe a
cross--like shape with the lower arm twisting eastwards. Compared to
Fig. \ref{surf} it is clear that small--number statistics may show
erroneous features and give a misleading image of the surface
distribution of a given type of stars. Though the selection of AGB
stars shown in Fig.  \ref{surf} has been more reliable than the
selection shown in Fig. \ref{surfjk}, the resulting sample is rather
incomplete when compared to the AGB stars selected from near--IR bands
only: once again it is very difficult to keep both reliability and
completeness at the required high level.

\subsubsection{The C/M ratio distribution}
\begin{figure*}
\centerline{%
\epsfxsize=0.5\hsize \epsfbox{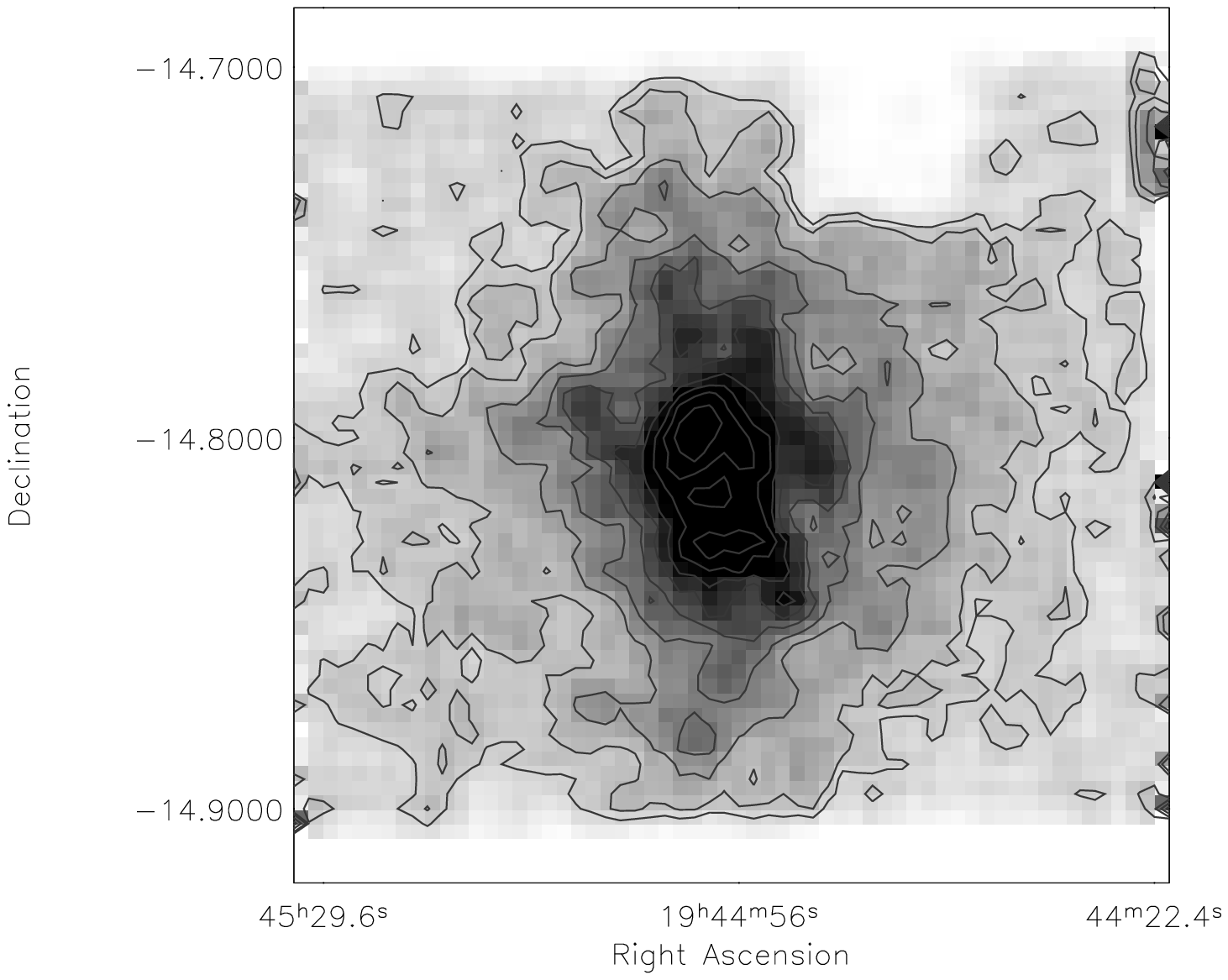}\quad
\epsfxsize=0.5\hsize \epsfbox{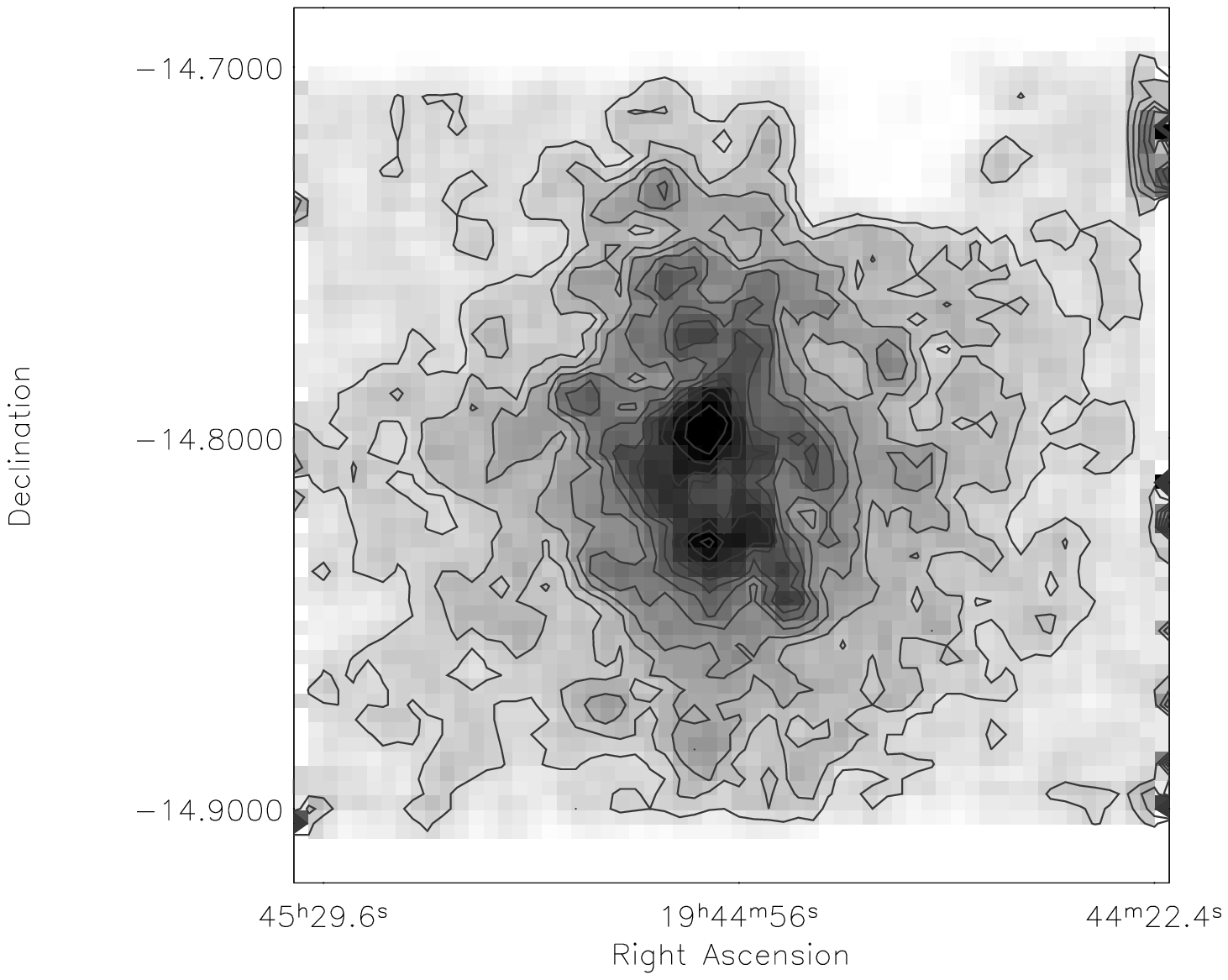}}
\smallskip
\centerline{%
\epsfxsize=0.5\hsize \epsfbox{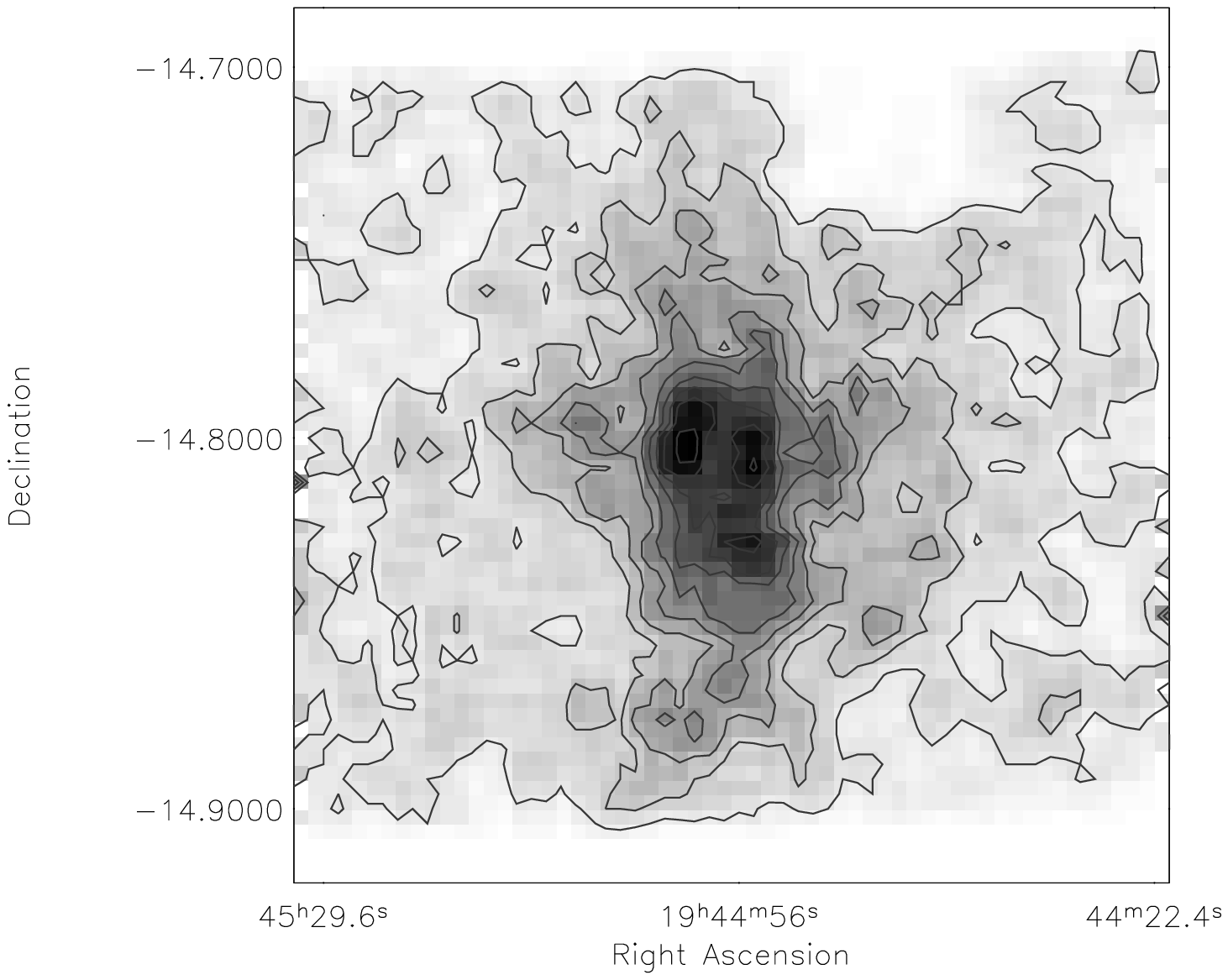}\quad
\epsfxsize=0.5\hsize \epsfbox{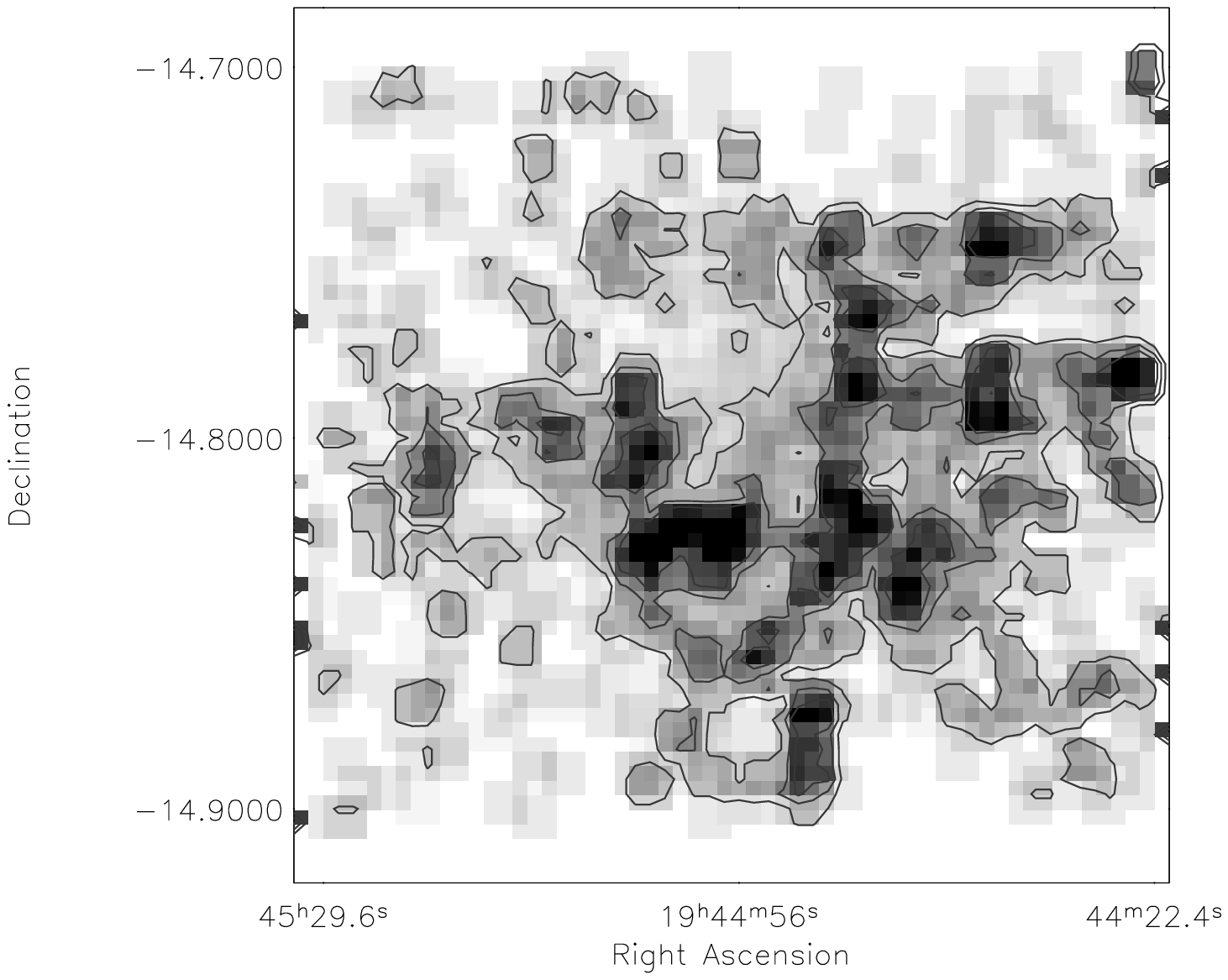}} 
\caption{Logarithmic and smoothed density distributions of stars selected
  only from $J$ and $K_{\mathrm{s}}$ data. Clockwise from top left, we show:
   all stars,  RGB stars,  AGB stars and  C/M ratio in
  $60\times 60$ bins of about $2^{\prime}$ square each. Darker regions
  correspond to higher density. Countours are: at $3$ and from $4$ to
  $20$ in steps of $=2=$ for all stars detected in $J$ and $K_{\mathrm
    s}$, from $1$ to $6$ in steps of $=1$ for RGB stars, from $1$ to
  $9$ in steps of $=1$ for AGB stars and from $0.15$ to $0.9$ in steps of 
  $=0.15$ for the C/M ratio distribution.}
\label{surfjk}
\end{figure*}

The distribution of the C/M ratio is similar to that in the Large
Magellanic Cloud (Cioni \& Habing \cite{cm}); it increases towards
some specific regions in the outer parts of the main optically visible
structure. The center has a rather low C/M ratio and this we consider to
be real (1.4); the low C/M ratio in the outer parts may be true but
the data are affected by small number statistics. This figure compared
to Fig. \ref{surf} describes almost exactly the same features, though
contours and grey scales are different because of the different
statistics.  Regions of higher C/M ratio are arranged in a broken
circle and avoid the region where the bar is located. The bar itself
has a relatively low C/M ratio in its northern parts, but a bar-like
structure cannot be recognized in the C/M distribution (e.g. Fig.
\ref{surfjk}). On the outer western side a secondary semi--circle is
seen that contains clumps with higher C/M ratio. This structure is parallel
to the inner broken--circle.  The two are separated by a region of lower
ratio values. Note a lack of stars in a square area in the NW; it is
purely artificial and due to a low quality image where too few sources
could be extracted. The largest elongated blob of high C/M ratio is
traceable in all other maps of Fig. \ref{surfjk} as the southern of two
components within the bar structure.

\begin{figure*}
\centerline{%
\epsfxsize=0.5\hsize \epsfbox{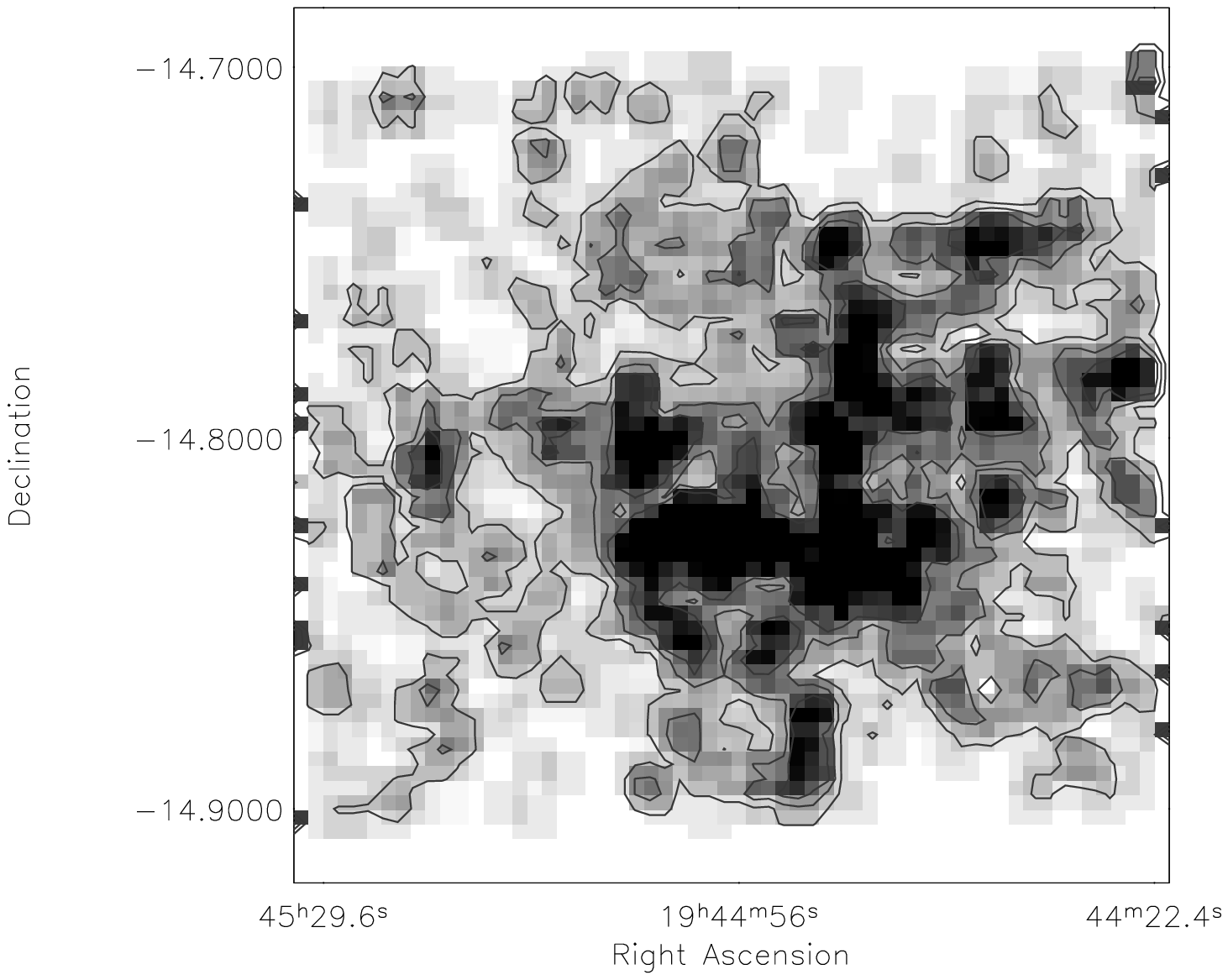}\quad
\epsfxsize=0.5\hsize \epsfbox{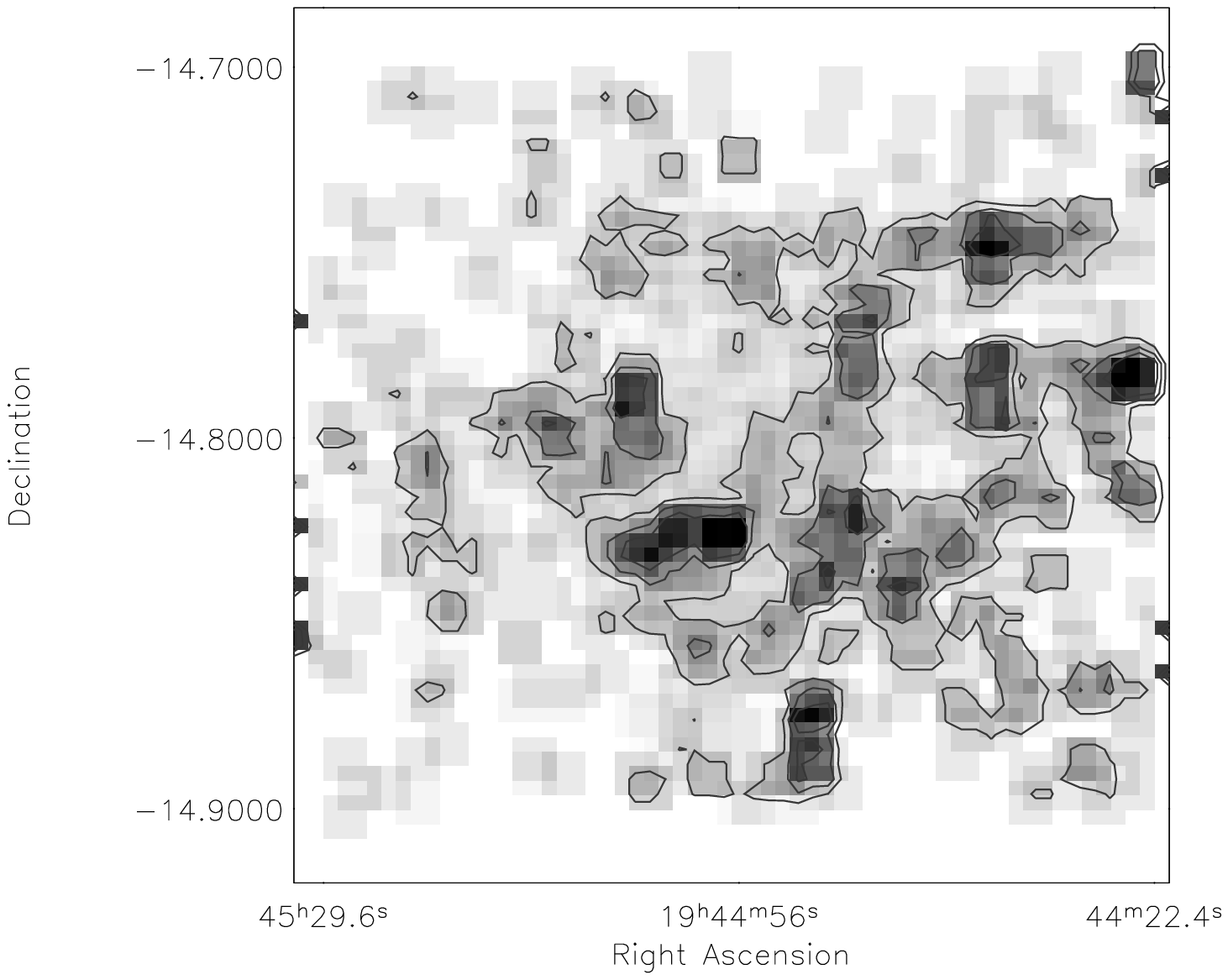}}
\caption{Logarithmic and smoothed density distributions of the C/M ratio 
  for stars selected only from $J$ and $K_{\mathrm{s}}$ data in  
  $60\times 60$ bins of about $2^{\prime}$ square each. C and M stars have 
been separated at $J-K_{\mathrm s}=1.26$ (left) and at $J-K_{\mathrm s}=1.46$ 
(right). Darker regions correspond to higher density. Countours are from 
$0.15$ to $0.9$ in steps of $=0.15$.}
\label{mapcm}
\end{figure*}

 The C/M  ratio is affected by the photometric errors-induced migration  
of stars through the colour  wall.  Figure  \ref{mapcm}
shows the distributions of the C/M ratio obtained by dividing AGB stars
into   C--rich  and   O--rich   at  $J-K_{\mathrm   s}=1.26$  and   at
$J-K_{\mathrm s}=1.46$.  Compared to the C/M  ratio distribution shown
in  Fig. \ref{surfjk},  where the  discrimination between  C--rich and
O--rich AGB stars  was set at $J-K_{\mathrm s}=1.36$,  the features of
high  ratio,  or  low  metallicity,  are  quite  similar.   The  major
difference is in the global C/M ratio. A variation of $0.1$ mag in the
selection criterium  of the two spectral  types of AGB  stars induces a
variation of about $0.15$ in the global C/M ratio.

Finally we would like to draw the attention of the reader to the absolute 
number of AGB stars involved in the C/M ratio distribution shown in 
Fig. \ref{surfjk}. The absolute number of C and M stars as a function of 
their C/M ratio is shown in Fig. \ref{num}. On the vertical axis integer 
numbers of C stars are indicated with triangles and of M stars with squares. 
The horizontal axis indicates the ratio between the number of C stars and M 
stars in a given ``column'', and it can be a fractional value. 
For example C/M$=0$ occurs because there are $0$ C stars in regions where 
there are from $1$ to $8$ M stars. C/M$=2$ occurs in 
a region where there are either $2$ C stars and $1$ M star, $4$ C stars and 
$2$ M stars or $6$ C stars and $3$ M stars. Though the absolute statistics 
are not striking, the occurrence of a high, or low, ratio in consecutive 
spatial regions makes it a more robust result.

\begin{figure}
\vspace{9cm}
\caption{Number of C stars (triangles) and M stars (squares) plotted versus 
their ratio.}
\label{num}
\end{figure}

\section{Discussion}

\subsection{Determination of the distance to the Galaxy}
The $I$--band magnitude of the TRGB depends only weakly on age and
metallicity (Salaris \& Cassisi  \cite{sal}) and can thus be used as
a standard candle; see Lee et al. (\cite{lee}). We adopt the interstellar
extinction as measured by Schlegel et al.  (\cite{sch}) of
$E(B-V)=0.24$, and the absorption in the $I$, $J$ and $K_{\mathrm s}$ wave 
bands available from NED derived using Cardelli's (\cite{cardi}) law. 
These are: $A_I=0.45$, $A_J=0.21$ and $A_K=0.08$. Combining the apparent
$I$--band TRGB magnitude derived in the previous section with the most
recent absolute calibration of the TRGB magnitude in the $I$--band by
Bellazzini et al.  (\cite{bel}), $M_I=-4.04\pm0.12$, we obtain
$(m-M)_0=23.34\pm0.12$; this corresponds to a distance of $466 \pm 10$
kpc.  Lee et al. (\cite{lee}) who also used the TRGB method measured
$(m-M)_0=23.46\pm0.10$.  Using $BVRI$ observations of Cepheids Gallart
et al.  (\cite{galla1}) derived $(m-M)_0=23.49\pm0.08$ while Clementini 
et al. (\cite{clem}) obtained $(m-M)_0=23.36\pm0.17$ from $BVI$ observations 
of RR Lyrae.  Our
measurement is, within the errors, in agreement with previously determined
values. The error we quote is only the formal error, but because we
adjusted the photometry to the DENIS $I$--band data we expect
systematic errors to be small. However, a major systematic uncertainty 
might be due to the absorption in the $I$ band. In fact Gallart et al. 
(\cite{galla1}) derive $I=19.8\pm 0.1$ while we obtain $I=19.76\pm 0.01$ which 
is in very good agreement. Note also that a large fraction of the uncertainty 
in the distance modulus comes from the error associated to the absolute 
calibration. Lee et al. (\cite{lee}) measure $I=20.05$ but derive a distance 
modulus very similar to that of 
Gallart et al. (\cite{galla1}). Beside the effect of 
a different extinction law there is also a difference in the method used to 
assign the TRGB location. Fig. 3 of Lee et al. (\cite{lee}) shows that for 
data grouped in bins of $0.1$ mag the convolution with an edge detection 
filter produces a histogram that peaks at $I=19.95$ with FWHM of about 
$0.3$ mag. The uncertainty is large and the number statistics of stars at 
the TRGB is about a factor of ten smaller than in our work and that of 
 Gallart et al. However there is an excellent agreement (within $0.02$ mag!) 
between our measurement and that derived from the mean luminosity of RR Lyrae 
by Clementini et al. (\cite{clem}).

\subsection{Metallicity gradient}

The significant variation in the C/M ratio over the face of the galaxy
is explained by a variation in the metallicity: a higher C/M implies a
lower metallicity as is well known from previous studies; for
references to earlier work on the SMC, LMC and the Milky Way galaxy
see Cioni\& Habing (\cite{cm}). The correlation between the C/M--ratio
and metallicity is qualitatively explained by Scalo \& Miller
(\cite{scalo}) and by Iben \& Renzini (\cite{ibre}): (i) O--rich AGB
stars of lower metallicity turn more easily into C--rich stars, (ii)
evolutionary tracks for lower metallicities are shifted to higher
temperatures and (iii) in very low metallicity environments
post--horizontal branch stars may fail to become AGB stars. If the
star formation rate has not varied with stellar mass, the variation in
C/M represents a variation in metallicity at the time that the
relevant AGB stars were formed. 
The variation in [Fe/H] over the face of NGC 6822 implies a lack of
circulation or turbulence: areas enriched in [Fe/H] in a local star
burst have not been mixed with areas of lower metallicity, at least
not within the age of these stars, between 2 and 5 Gyr.

\begin{figure}
\resizebox{\hsize}{!}{\includegraphics{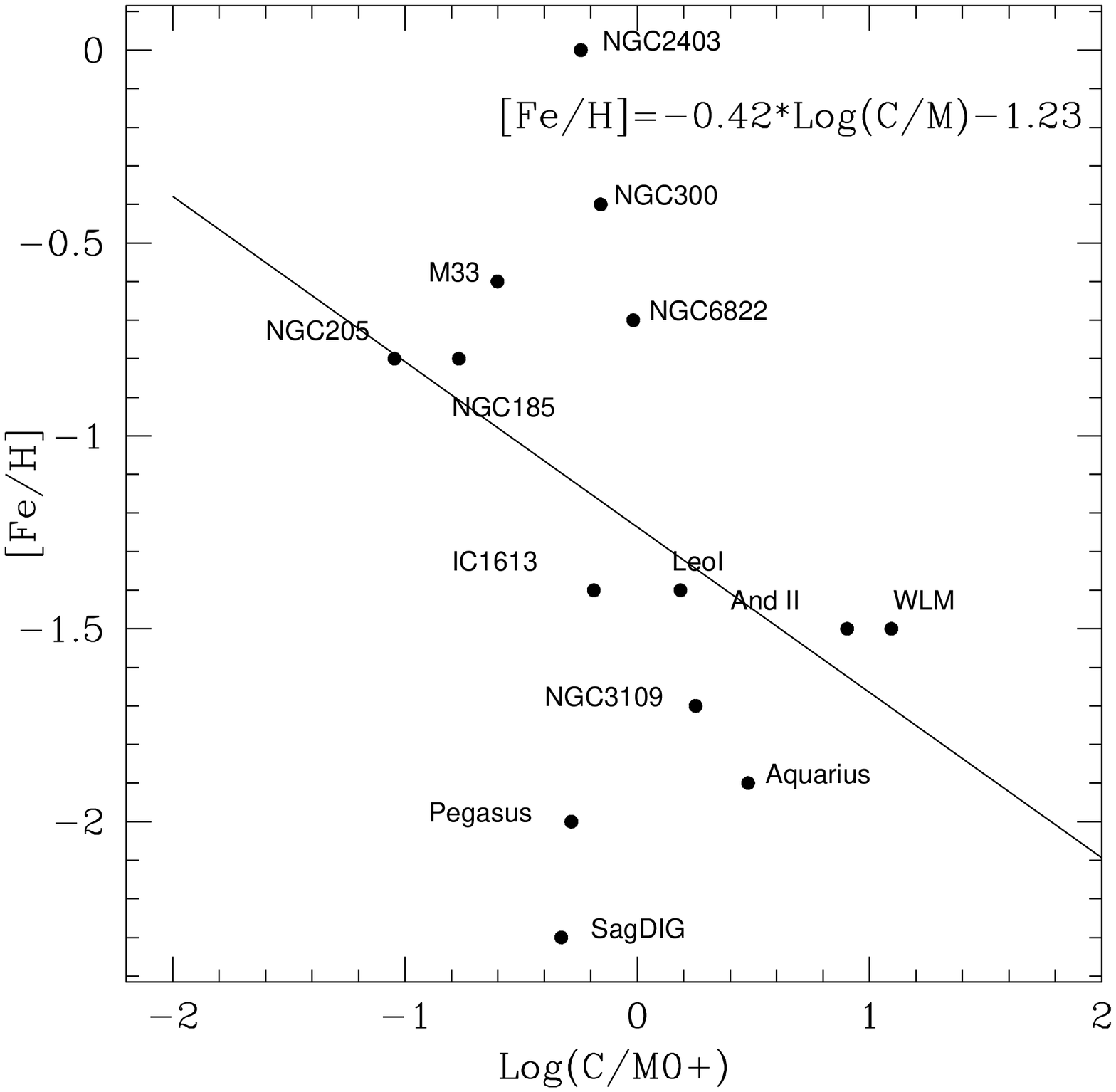}}
\caption{Relation between the metallicity ([Fe/H]) and the decimal logarithm 
of the C/M ratio for M0+ stars. Data are taken from Groenewegen 
(\cite{gronew}). The solid line is a linear fit through the data.}
\label{newr}
\end{figure}

A calibration of the relation between C/M and [Fe/H] is presented in
Cioni \& Habing (\cite{cm}) : $[Fe/H]=-1.08\times Log(C/M)-1.05$.
This relation is rather uncertain (about $0.2$ dex in each
[Fe/H] measurement and about $50$\% in the total number of AGB stars
detected in each galaxy) but it still provides a good indication of the
spread in metallicity. We have updated this relation using the latest 
number of M0+, C and [Fe/H] from Groenewegen (\cite{gronew}) 
and by correcting a typo in Groenewegen's Table 1.1, kindly pointed out by 
the referee P. Battinelli: 
the number of M0+ stars in SagDIG is $34$ instead of $1$. We obtain: 
$[Fe/H]=-0.42\times Log(C/M)-1.23$ as shown in Fig. \ref{newr}. 
The relation becomes less steep 
because the effect of SagDIG is compensated by NGC 2403. Though the latter is 
outside the Local Group boundary (1 Mpc) there is no apriori reason to 
exclude it.

C/M contours in Fig. \ref{surf} and Fig.
\ref{surfjk} vary from $0.3$ to $0.9$ and $0.15$ and $0.9$
respectively and this then corresponds to a variation of $0.25$ dex
and $0.31$ dex in metallicity respectively.  However, the absolute and
unsmoothed variation of C/M is about $6$ and this corresponds to an
average $\Delta$[Fe/H] $=-1.56$ dex.  This spread is about twice 
the value derived for both Magellanic Clouds and agrees with the
spread derived by Tolstoy et al. (\cite{tolto}). The authors measured
the Ca II triplet in $23$ RGB stars and obtained on average
[Fe/H]$=-1.0\pm0.5$ but with the highest metallicity around
[Fe/H]$=-0.5$ and the lowest around [Fe/H]$=-2.0$. Instead a gradient
in [O/H] has been suggested by Venn et al. (\cite{venn}).  At present
we are measuring the Ca II triplet absorption lines to check and
perhaps improve the metallicity index of a statistically significant
number of AGB stars in NGC 6822.

\subsection{The structure of NGC 6822}

The near--IR data covering the whole galaxy allow us to investigate
its structure and derive viewing angles: the inclination ($i$) and
position angle (PA) of the line of nodes (i.e. the intersection of the
galaxy plane and the sky plane) following a technique developed by van
der Marel \& Cioni (\cite{vdm}).  We have transformed sky coordinates
($\alpha$, $\delta$) of sources in {\it table2.dat} into angular
coordinates ($\rho$, $\phi$) for all sources detected in $J$ and
$K_{\mathrm s}$ regardless of their $I$--band values. We thus cover NGC 6822
entirely while the sources remain well resolved even in the most
crowded central region. We use as central reference the point
($\alpha=19$:$44$:$56$, $\delta=-14$:$48$:$06$) that corresponds to
the optical and radio center of the galaxy (van den Berg \cite{vdb}).
The angle $\phi$ is measured counterclockwise starting from the West.
We divided the surface of NGC 6822 into $8$ sectors, each with an
opening angle of $45$ deg. and into $2$ rings: $\rho<0.06\deg$ and
$0.06\deg<\rho<0.12\deg$. This division provides
a minimum significant number of stars in each region; there are in
total $16$ separate regions.  The goal is to investigate how much the
magnitude of a group of stars varies with $\rho$ and $\phi$. Thus we
selected C--rich AGB stars with $1.4<(J-K_{\mathrm s})<2.0$ and
$15<K_{\mathrm s}<16$ and calculated their number distribution in
colour and magnitude in each region.  Assuming that the stars that we
see in the galaxy lie in a thin plane, the magnitude at a given distance
from the center will have an approximately sinusoidal variation with
position angle with amplitude $A=0.038\rho\ tan\, i$.

\begin{figure}
\resizebox{\hsize}{!}{\includegraphics{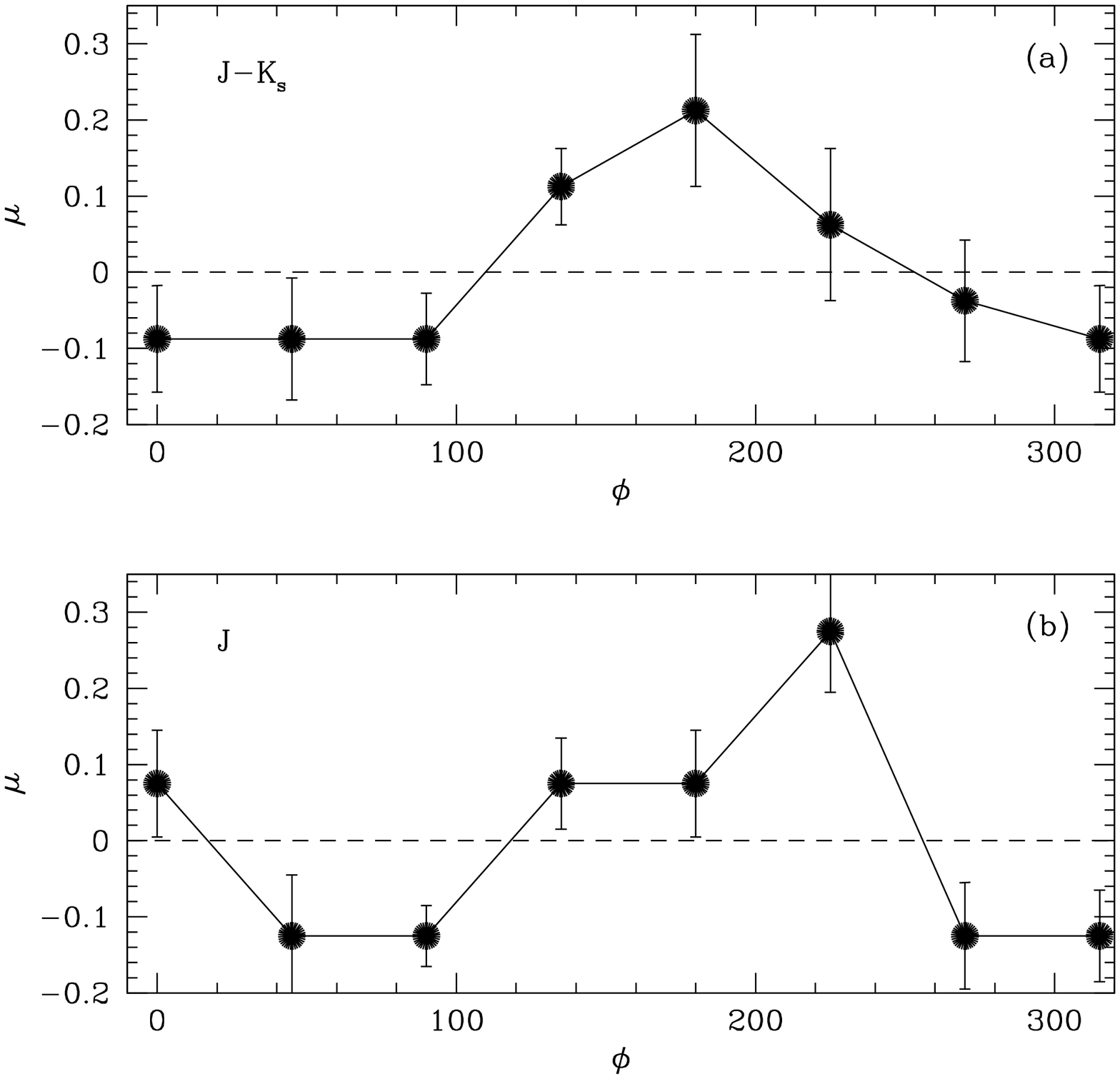}}
\caption{Difference between the magnitude of the mode of the $J-K_{\mathrm s}$ 
colour (a) and $J$ mag (b) of C--rich AGB stars distributed  
in $8$ separate sectors of a ring between $0.06$ and $0.12$ deg from the 
center and their mean. Errors equal $\sigma /\sqrt{N}$.}
\label{stru}
\end{figure}

The major limitation in applying this technique in NGC 6822 lies in
the low numbers of stars. We found a systematic variation in the ring
at $\rho=0.09$ deg but not in the other ring; see Fig. \ref{stru}.
From this we derive the inclination, $i\approx 89$ deg. This result is
considerably different from the value deduced from Letarte et al.
(\cite{tarte}): taking the ellipticity of their outer contour, $e=0.1$
and using the formula $i=arccos (1-e)$ we obtain an inclination angle
of only $i=25$ deg.  If $i=25$ deg is the real inclination of the
galaxy then we expect an amplitude variation, as defined above, of the
order of $10^{-3}$ mag, too small to explain the variation observed in
this one ring. Note that the formal error associated to each point in
Fig. \ref{stru} is a factor $\sqrt N$ smaller than the observed variation
observed which is on the other hand comparable to the formal error
associated to each single point.  The intersection between the
continuous line and the zero level line corresponds to the PA defined as
PA$=\phi-90$ in the usual astronomical convention (i.e. measured
counterclockwise starting from the North), therefore we obtain
PA$\approx10$ deg which is in very good agreement with the PA of the
optical bar derived by Hodge (\cite{hodge}).

It is interesting to note that also the $J-K_{\mathrm s}$ colour has
an almost regular variation as a function of position angle. This may
be due to a variation in metallicity; if so, it should correspond to
the variation in metallicity derived from the C/M ratio. In fact
comparing with Fig. \ref{surf} and \ref{surfjk}, sources in the outer
NE and SE regions have a lower C/M ratio than sources in the outer NW
and SW regions.

The situation is more confused in the inner ring and no regular
pattern could be detected for the $K_{\mathrm s}$ band either.
The latter is expected because C--rich AGB stars span a larger range
in $K_{\mathrm s}$ than in $J$, thus the $J$ band is by definition a
much better indicator of differences in distance unless a very large
number of stars is present.

Finally we did not study the variation of the TRGB as a function of sector 
in the outer ring. This should be done using $I$--band data as well, but the 
effect of the central gap, crowding and seeing would strongly bias the 
result.

\subsection{Comparison with other distributions}
Gallart et al. (\cite{galla3}) found that recent star formation
($100$--$200$ Myr) is enhanced in the upper and lower part of the NGC
6822 bar compared to the center, and this has been confirmed recently by de
Blok \& Walter (\cite{blok3}).  Maps in Fig. \ref{surfjk}, and in
particular the distribution of RGB stars, show a high central density
of stars that would be on average several Gyr old, in agreement with
Gallart et al. (\cite{galla3}). The central region is also populated
by a high concentration of AGB stars as well as the lower part of
the bar. Deeper photometric observations obtained by Wyder
(\cite{wyde}) with the Hubble Space Telescope also show an old RGB
population in the central bar region. This region has also
experienced, like other fields surveyed in the bar, a higher rate of
star formation in the past $0.6$ Gyr.  The high extinction of this
region is in agreement with a high $60\mu$m emission as measured by
the IRAS satellite (Rice \cite{rice}).

Letarte et al. (\cite{tarte}) have surveyed NGC 6822 in four filters,
two broad--band filters and two narrow--band filters, to identify and
study the distribution of C stars -- a technique first used by Cook et
al. (\cite{cook}).  They detected $904$ C stars and several RGB stars
that describe an almost elliptical structure as far out as $\rho=
23.7^{\prime}$. In this study we select $500$ C stars using the
combined $IJK_{\mathrm s}$ criteria and $1511$ using $JK_{\mathrm s}$
photometry only.  The elliptical structure claimed by Letarte is
dominated by RGB stars (all their stars with $I<21.7$) and this is
directly comparable with the distribution of the sources that we
selected.  Figures \ref{surfjk} and \ref{surf} show that the sources are
widely distributed and that a slightly elliptical structure is
confirmed; most importantly this broad structure is not only formed by
old stars.  The extended distribution of RGB stars in Fig. \ref{surf}
is only marginally afftected by foreground sources because of the
adopted selection criteria.  This permitted us to measure the C/M
ratio across the galaxy, something that Letarte et al. (\cite{tarte})
did not explore.

The distribution of HI gas (de Blok \& Walter \cite{blok}, Weldrake et
al.  \cite{weld}) extends to $\approx \pm 40^{\prime}$ and defines an
elongated, flattened structure with PA$=125$ deg. and $i\approx
60$ deg that is rather clumpy, especially in the center. There is an
indication that some of the high C/M ratio blobs correspond to high HI
column density regions.  The distribution of the HI gas usually
indicates regions where star formation is currently taking place. We
may conclude that where the correspondence holds the gas is relatively
metal poor. Mouhcine \& Lan\c{c}on (\cite{mou}) suggest that the C/M
ratio measures both the metallicity of the carbon star progenitors and
the present interstellar medium metallicity. Though we cannot exclude
that the local composition has been altered by the presence of a bar
we certainly know that because NGC 6822 is isolated within the Local
Group, it has not suffered strong interactions that may have altered its
inner disk.  A more detailed comparison between the two distributions
and follow--up studies of the stellar populations associated to the
different features will put further constraints on the evolution of the
galaxy.

\subsection{Miscellaneous}

The average metallicity of NGC 6822 can be derived from the dereddened
$(J-K_s)_0$ colour using the relation(s) by Valenti et al.
(\cite{val}). Based on near--infrared data of several globular
clusters in the Galaxy these authors calibrated the photometric
indices that describe the location and morphology of the RGB. Here we
refer to the $(J-K_s)_0$ index and we will use only the relation
$(J-K)_0=0.22\times $[Fe/H]$+1.14$ at $M_K=-5.5$. Applying the
reddening correction and the distance modulus as determined in the
previous section we obtain a CMD of $(J-K)$ versus the absolute
magnitude $M_K$; see Fig.  \ref{jkabs}. The box in the
upper right corner shows a histogram of all sources within $0.01$ mag
from $M_{K}= -5.5$ as a function of $(J-K_s)_0$. A well
defined peak is present at $(J-K_s)_0=0.9$ from which we derive a
metallicity of [Fe/H]$=-1.09\pm 0.2$; here we have assumed that the
uncertainty of this determination is represented by half the bin size
($0.1$ mag). This value agrees with a value derived by Davidge
(\cite{dav}) from the slope of the RGB -- a quantity that we cannot
measure because our observations are not deep enough. Because the
variation in the C/M ratio suggests a considerable spread in
metallicity, one can speculate that the metallicity obtained from RGB
and AGB stars relates to two different epochs of star formation; in a
simple stellar population the RGB stars are on average older than AGB
stars. If the age difference between AGB and RGB stars is of the order
of $2$ Gyr and the chemical enrichment law according to Davidge is 
$\Delta$[Fe/H]$\Delta t=-0.2\pm 0.1$ dex Gyr$^{-1}$, we conclude that
an average AGB star in NGC 6822 would have [Fe/H]$=-0.89$ dex.

\begin{figure}
\vspace{9cm}
\caption{Colour--magnitude diagram ($(J-K_s)_0$, $M_K$) of sources
  detected in NGC 6822 in $J$ and $K_s$. The histogram in the corner
  shows the number of source as a function of colour within $0.01$ mag
  around $M_K=-5.5$. The colour corresponding to the peak is a measure
  of the average galaxy metallicity.}
\label{jkabs}
\end{figure}

The bolometric magnitude of AGB stars in NGC 6822 can be qualitatively
derived as follows. AGB stars shown in Fig. \ref{jkk} are likely to
have thin circumstellar envelopes and the $I$, $J$ and $K_s$ wave
bands will contain most of the stellar flux. These AGB stars have on
average $I=19.2$ and $(I-J)=1.5$. Using the relation derived by
Alvarez et al. (\cite{alva}), $m_\mathrm{bol}-I=1.32-0.574\times
(I-J)-0.0646\times(I-J)^2$ -- this relation has been compared with the
integration over the spectral energy distribution of AGB stars in the
Small Magellanic Cloud observed from the optical $I$--band to the
mid--IR $LW10=12\mu$m ISO--band (cft. Appendix in Cioni et al.
\cite{iso}), correcting for extinction and by applying the distance
modulus derived in Sect. 5.1, we obtain that the average AGB star in NGC
6822 would have $M_\mathrm{bol}=-4.3$.  Furthermore, assuming that at least a
few stars with these characteristics are typical Mira variables and
pulsate with $P\approx 250$ days we can also use the theoretical
models by Vassiliadis \& Wood (\cite{vassi}), that relate $M_\mathrm{bol}$
with $Log(P)$ to derive that the average AGB star in NGC 6822 has a
main-sequence mass of about $1.5$ M$_\odot$ and is about $2$ Gyr old.
These are purely qualitative considerations and have been discussed
only to obtain a crude indication of the type of AGB stars detected in
this paper. Only when accurate measurements of the pulsation period
are available will we put more quantitave constraints on the mass and
age of the AGB stars. A monitoring program is currently under way.

\section{Conclusions}
In this paper we report observations in the $I$, $J$ and
$K_{\mathrm s}$ bands of the central $20^{\prime}\times 20^{\prime}$
of NGC 6822. Many RGB and AGB stars have been detected and their
distribution in colour--colour and colour--magnitude diagrams has been
discussed. In particular RGB stars and AGB stars (both C--rich and
O--rich) have been statistically disentangled. The surface
distribution of these groups and of the C/M ratio, as an indicator of
metallicity, have been discussed. We derive a  spread in
metallicity within NGC 6822 that is twice as large as the spread deduced
within each Magellanic Cloud and that corresponds to about $1.56$
dex. The average metallicity of an average AGB star of $1.5$
M$_\odot$, approximately $2$ Gyr old, detected in this study, could be
$-0.89$ dex. Regions of higher ratio, or of lower metallicity, are
distributed in one or two semi--circles around the location of the
bar which itself has a lower ratio or a higher metallicity. The
position of the TRGB in the $I$ band has been used to derive a
distance modulus for the galaxy of $(m-M)_0=23.34\pm0.12$.  The
photometry discussed in this paper has been
calibrated using DENIS data for the $I$ band, and 2MASS data for the $J$
and $K_{\mathrm s}$ bands. C--rich AGB stars distributed in a ring
between $0.06$ and $0.12\deg$. from the galaxy center show a different
$J$--band mode magnitude as a function of position angle. The
amplitude of this difference corresponds to an inclination of about
$89$ deg and a PA$\approx 10$ degrees. A similar behaviour is found
for the $J-K_{\mathrm s}$ colour that is probably a result of a
difference in metallicity.  If the HI gas distribution traces the
plane of the galaxy then the small bar may be the result of 
projection of a more extended bar--like structure viewed at an
inclination of about $89$ degrees.

\begin{acknowledgements}
We  thank  Philip Habing  for  his  enthusiastic  contribution to  the
observations and the referee Paolo Battinelli for his useful comments 
that enhanced the analysis and discussion presented in this paper.   
 This paper  makes use  of data  products from  the Two
Micron All Sky  Survey, which is a joint project  of the University of
Massachusetts    and   the    Infrared    Processing   and    Analysis
center/California  Institute  of Technology,  funded  by the  National
Aeronautics  and   Space  Administration  and   the  Mational  Science
Foundation. The paper also uses data from the DENIS project that is partially
funded by  European Commission through  SCIENCE and Human  Capital and
Mobility plan grants, and is  also supported, in France by the Institut
National  des Sciences de  l'Univers, the  Education Ministry  and the
Centre National de la Recherche  Scientifique, in Germany by the State
of  Baden-W\"urtemberg,  in Spain  by  the  DG1CYT,  in Italy  by  the
Consiglio  Nazionale  delle Ricerche,  in  Austria  by  the Fonds  zur
F\"orderung  der  wissenschaftlichen  Forschung und  Bundesministerium
f\"ur Wissenschaft  und Forschung,  in Brazil by  the Foundation
for the development  of Scientific Research of the  State of Sao Paulo
(FAPESP), and in Hungary by an OTKA grant and an ESOC\&EE grant.  This
research has made use  of NASA/IPAC Extragalactic Database (NED) which
is operated by the  Jet Propulsion Laboratory, California Institute of
Technology,  under  contract   with  National  Aeronautics  and  Space
Administration.
\end{acknowledgements}

\end{document}